\documentclass[prb,twocolumn,showpacs,preprintnumbers,amsmath,amssymb,floatfix]{revtex4}

\usepackage{graphicx}
\usepackage{dcolumn}
\usepackage{bm}

\begin{document}

\title{\bf Spectral properties of the dimerized and frustrated $S=1/2$ chain} 

\author{Kai P. Schmidt, Christian Knetter and G\"otz S. Uhrig}

\affiliation{Institut f\"ur Theoretische Physik, Universit\"at zu
  K\"oln, Z\"ulpicher Strasse 77, D-50937 K\"oln, Germany\\
  {\rm(\today)} }

\begin{abstract}
Spectral densities are calculated for the dimerized and frustrated 
$S=\frac{1}{2}$ chain using the method of continuous unitary transformations 
(CUTs). The transformation to an effective triplon model is realized in a perturbative fashion up to high orders about the limit of isolated dimers. An efficient description in terms of triplons (elementary triplets) is possible: a detailed analysis of the 
spectral densities is provided for strong and intermediate dimerization including the
 influence of frustration. Precise predictions are made for inelastic neutron scattering experiments probing the $S=1$ sector and for optical experiments (Raman scattering, infrared absorption) probing the $S=0$ sector. Bound states and resonances influence the important continua strongly. The comparison with the field theoretic results reveals that the sine-Gordon model describes the low-energy features for strong to intermediate dimerization only at critical frustration. 
\end{abstract}

\vskip2pc

\maketitle
\section{Introduction}
One-dimensional quantum antiferromagnets display very interesting and  
fascinating physical properties. The generic system to be studied in this 
context is the dimerized and frustrated Heisenberg chain which comprises 
many physical phenomena of interest. There are gapless and 
gapped phases\cite{halda82b}, fractional excitations\cite{cloiz62,fadde81} and 
confinement\cite{halda82b}.

The quantitative calculation of spectral densities is a very important issue 
in the field of strongly correlated electron systems. 
The interplay between kinetics, interaction and 
matrix elements leads to characteristic structures in the spectra. In 
recent years there has been significant progress in calculating spectral 
densities in the field of quasi one-dimensional quantum spin systems, 
e.g., the uniform Heisenberg chain\cite{karba97}, spin ladders systems
\cite{knett01b,schmi01} and strongly dimerized spin chains
\cite{zheng03a,hamer03}.

Beside the fascinating theoretical aspects, the determination of spectral 
densities is of direct importance for experimental measurements. 
The theoretically predicted spectra are relevant for inelastic neutron 
scattering experiments and optical experiments like Raman spectroscopy and 
infrared absorption. There is a large number of quasi one-dimensional 
compounds which can be successfully described by the dimerized and frustrated 
Heisenberg model, e.g., spin-Peierls compounds like CuGeO$_3$
\cite{hase93a,nishi94,riera95,casti95} and $\alpha^\prime$-NaV$_2$O$_5$
\cite{isobe96}, (VO)$_2$P$_2$O$_7$\cite{garre97a} or organic compounds like 
Cu$_2$(C$_5$H$_{12}$N$_2$)$_2$Cl$_4$\cite{chabo97a} and 
(Cu(NO$_3$)$_2\cdot 2.5$D$_2$O)\cite{xu00,tenna03}, but also undimerized 
cuprate chain compounds like KCuF$_3$\cite{nagle91}, 
Sr$_2$CuO$_3$\cite{ami95} and SrCuO$_2$\cite{motoy96}.
 
We will describe the dimerized and frustrated spin chain in terms of 
elementary triplets (triplons)\cite{notiz1} which carry total spin one. The 
commonly accepted elementary excitations for the undimerized system are 
fractional excitations, so-called spinons carrying 
$S=1/2$\cite{cloiz62,fadde81}. Recently we have shown 
that a description in terms of triplons is also
possible for the isotropic Heisenberg chain\cite{schmi03c}. So 
there is no necessity to use
 fractional excitations in one-dimensional systems.
Remarkably, even more spectral weight is captured by the states of
two triplons than with two spinons\cite{schmi03c}. This result was recently confirmed by Hamer {\it et al}.\cite{hamer03}.

 In this work, we take a closer look at the generic features of the spectral 
properties at finite dimerization. In this regime a description in terms of 
triplons is surely correct due to the confinement of the 
spinons\cite{halda82b}. We will learn more about the triplon
 picture and try to connect the triplon-based findings at finite dimerization
 with results obtained at zero dimerization using the spinon picture.
  
The paper is organized as follows. Sect.~II gives an introduction in the model
 we consider in this work. Sect.~III presents the method we use and introduces 
the basic quantities under consideration. In Sect.~IV results for the dynamical
 structure factor are shown. We present results for the one-triplon 
contribution, the two-triplon contribution and compare  with
field theoretical results. Sect.~V shows spectral densities with total spin zero. 
We give results for the two-triplon contribution for nearest neighbor (NN) 
and next-nearest neighbor (NNN) coupling. We also provide results for Raman
 spectroscopy and optical absorption. Sect.~VI summarizes this paper and 
Sect.~VII comprises the conclusions.
    
\section{Model} 
The Hamiltonian for the dimerized and frustrated ${\bf S}=1/2$ spin chain reads
\begin{equation}
 \label{H_DFC}
 H = J_0\sum_i \left[ \left(1+\delta(-1)^i\right){\bf S}_i{\bf S}_{i+1} +
 \alpha_0 {\bf S}_i {\bf S}_{i+2} \right],
\end{equation}
where $\delta$ parameterizes the dimerization and $\alpha_0$ the relative
 frustration between next-nearest neighbor spins. In order to apply a 
perturbative treatment we transform Eq.~(\ref{H_DFC}) into
\begin{equation}
 \label{H_TDFC}
 H/J =  \sum_i \left[ {\bf S}_{2i}{\bf S}_{2i+1} + 
\lambda{\bf S}_{2i}{\bf S}_{2i-1} + \lambda \alpha {\bf S}_{i}
 {\bf S}_{i+2} \right],
\end{equation}
where $J=J_0(1+\delta)$, $\lambda=(1-\delta)/(1+\delta)$ and 
$\alpha=\alpha_0/(1-\delta)$.

\begin{figure}
  \begin{center}
    \includegraphics[width=\columnwidth]{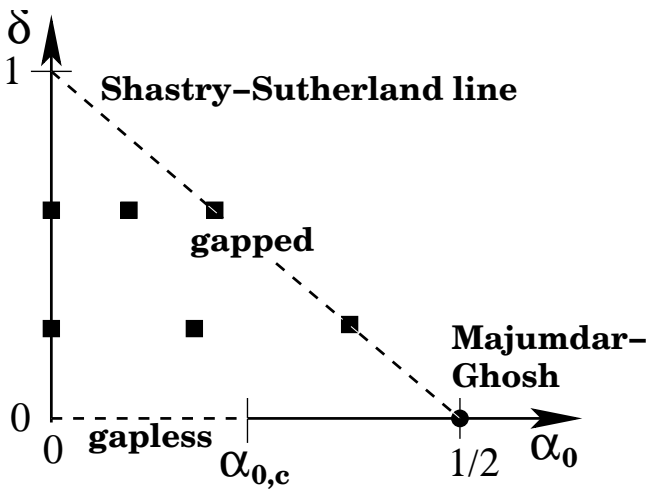} 
    \caption{Phase diagram of the dimerized and frustrated Heisenberg chain
 depending on frustration $\alpha_0$ and dimerization $\delta$. The system is
 always in a gapped regime except in the interval 
$\alpha_0\in[0,\alpha_{0,c}]$ at zero dimerization. The dashed line marks the
 Shastry-Sutherland line where the ground-state is known exactly. Solid squares correspond to the parameters $(1-\delta)/(1+\delta)=\lambda\in\{0.3;0.6\}$ and $\alpha_0/(1-\delta)=\alpha\in\{0.0;0.25;0.5\}$ used in this work.} 
    \label{fig:Phase}
  \end{center}
\end{figure}
The dimerized and frustrated spin chain exhibits very interesting intrinsic
 physics. The phase diagram of the model is shown in Fig.~\ref{fig:Phase}.
 At $\delta=0$ there are two regimes. 

(i) For $\alpha_0<\alpha_{0,c}$ the ground 
state is in the same universality class as the uniform Heisenberg chain. 
The excitations are massless and the standard description is in terms of 
unconfined spinons carrying total spin $S=1/2$\cite{cloiz62,fadde81}. In a recent work 
we have shown that also a description in terms of triplons is appropriate for
 the isotropic Heisenberg chain\cite{schmi03c}.

(ii) At $\alpha_0=\alpha_{0,c}$ there is a transition into a spontaneously 
dimerized phase\cite{casti95,julli83,okamo92,egger96}. The ground-state is 
two-fold degenerate and the excitations are massive spinons for 
$\alpha_0>\alpha_{0,c}$. At the Majumdar-Ghosh point ($\alpha_0=0.5$) the 
ground-state is known exactly\cite{majum69a,majum69b,broek80}. The validity 
of a triplonic description in this gapped phase is an open issue. On the other
 hand, Haldane\cite{halda82b} has shown that for any finite dimerization 
$\delta$ the spinons become confined and a description in terms of triplons 
carrying total spin $S=1$ is appropriate. The spectrum is always gapped
\cite{halda82b,shast81a,chitr95} and the excitations can be viewed as bound 
states of two spinons\cite{luthe75,halda80,uhrig96b,gogol98}. The interaction 
between the spinons is not exhausted by the confinement and there remains
a triplon-triplon interaction which can lead to two-triplon bound 
states with total spin $S=0$ and $S=1$ lying below the multi-triplon continuum
\cite{shast81a,uhrig96b,affle97,fledd97,bouze98a,shevc99,barne99,trebs00,zheng01a}.

Spectral properties are particularly difficult to address. So far,
results can be obtained either by numerical approaches like exact diagonalization
or quantum Monte Carlo or by studies of effective continuum models.
In particular, the case  of finite frustration and the $S=0$ sector relevant for 
optical experiments has not yet been investigated thoroughly.

In the following we expand about the limit of isolated dimers on the strong 
bonds, i.e $\lambda=0$. We present results for the two-triplon
contribution to the spectral density for strong ($\lambda=0.3$) and
intermediate ($\lambda=0.6$) dimerization and for three representative
values of the frustration ($\alpha=0$,$\alpha=0.25$,$\alpha=0.5$). These parameters are marked in Fig.~\ref{fig:Phase} as filled squares. 

\section{Method}

A continuous unitary transformation\cite{wegne94} is used to map the 
Hamiltonian $H$ to an effective Hamiltonian $H_{\rm eff}$ which conserves the
 number of triplons on the strong bonds, i.e.\ $[H_{\rm 0},H_{\rm eff}]=0$ 
where $H_{\rm 0}:=H|_{\lambda=0}$ \cite{knett00a,knett03a}. The ground state
 of $H_{\rm eff}$ is the triplon vacuum. Using an infinitesimal antihermitean
 generator $\eta$ we have  
\begin{equation}
 \label{CUT}
 \frac{dH}{dl} = [\eta (l),H(l)]\,
\end{equation} 
where $H(0) = H$, $H(\infty) = H_{\rm eff}$ and $l$ is an auxiliary variable.
 The optimized choice for $\eta$ reads
\begin{equation}
 \label{Generator}
 \eta_{i,j}(l) = {\rm sgn} ([H_{0}]_{i,i}-[H_{0}]_{j,j})H_{i,j}(l)
\end{equation}
where the matrix elements $\eta_{i,j}$ and $H_{i,j}$ are given in an 
eigen-basis of $H_{\rm 0}$ counting the number of triplons on the strong 
bonds. The choice (\ref{Generator}) retains only triplon conserving processes
and it eliminates all parts of $H$ changing the number of triplons\cite{knett00a}.
It must be stressed that the approach represents a renormalization in the sense
that matrix elements between states with very different energies are transformed
strongest. The matrix elements between energetically similar states are transformed
only at a later stage of the transformation.
In order to determine spectral weights $I_{\rm n}$ and spectral densities the 
observable $R$ is mapped to an effective observable $R^{\rm eff}$ by the same unitary transformation\cite{knett03a} 
(\ref{CUT}) as the Hamiltonian. 
 
The continuous unitary transformation cannot be carried out without truncation.
We use a perturbative method in $\lambda$. The effective hamiltonian $H^{\rm eff}$
is calculated up to order $10$ and the effective observable $R^{\rm eff}$ is 
calculated up to order $7$ in the two-triplon sector. The plain series of the important quantities will be made available on our home pages\cite{home}. The numerical effort to obtain the result is discussed in detail for the spin-ladder systems and will not be discussed in this work\cite{knett03b}.

The following extrapolation technique is employed which we have introduced
 recently for spin-ladder systems\cite{schmi03a,schmi03d}. After fixing 
$\alpha$ to the desired value the plain series in $\lambda$ is converted into
a series in $1-\Delta$ invoking the one-triplon gap. 
The one-triplon gap is the natural internal 
energy scale of the problem. Since in this work we are interested only in 
strong and intermediate dimerization, no further extrapolation techniques 
like standard Pad\'{e} extrapolants are used. There is no uncertainty in the 
obtained spectral densities for strong dimerization $\lambda=0.3$. The uncertainty is  
about $2\%$  for the worst case at intermediate dimerization 
$(\lambda=0.6)$. In order to investigate the cases of weak or vanishing 
dimerization it would be important to use further extrapolation tools and to
 treat processes with longer or infinite range explicitly.

Spectral densities can be calculated at $T=0$ by the retarded resolvent
\begin{equation}
 \label{Intensity}
 I(\omega) = -\frac{1}{\pi}{\rm Im}
\left\langle0\left|R^{\dagger}\frac{1}{\omega-(H-E_0)+ i0+}R\right|0\right\rangle \ .
\end{equation}
Due to the conservation of triplons after the unitary transformation the 
spectral density can be split into additive parts $I_n(\omega)$, the 
$n$-triplon contribution to the spectral density,
\begin{eqnarray}
 I(\omega) &=& \sum_n I_n (\omega) \\\nonumber
           &=& -\frac{1}{\pi}{\rm Im}\left\langle0\left|R_{\rm eff,n}^{\dagger}
\frac{1}{\omega-H_{\rm eff,n}+i0+}R_{\rm eff,n}\right|0\right\rangle \ .
\end{eqnarray}
The total intensity can be calculated using Dirac's identity for 
Eq.~(\ref{Intensity})
\begin{equation}
 I_{\rm tot} = \int_{0+}^{\infty} d\omega I(\omega) = 
\langle0|R^{\dagger,{\rm eff}}R^{\rm eff}|0\rangle \ .
\end{equation}
The total intensity $I_{\rm tot}$ is a sum over the spectral weight of all
 triplon sectors, $I_{\rm tot}=\sum_{\rm n}I_n$. The spectral weight 
$I_{\rm n}$ in the ${\rm n}$-triplon sector is calculated by 
\begin{equation}
 \label{Itot}
 I_{\rm n} = \langle0|R^{\dagger,{\rm eff}}_{\rm n}R^{\rm eff}_{\rm n}
 |0\rangle 
\end{equation} 
where $R_{\rm n}$ denotes all $n$-triplon excitation processes of the 
observable $R$. Using the sum rule 
$I_{\rm tot}=\langle0|R^2|0\rangle-\langle0|R|0\rangle^2$ we can check the
reliability of the perturbative results. For later use we define 
the relative spectral weights $I_{\rm n, rel}=I_{\rm n}/I_{\rm tot}$ with
 $\sum_n I_{\rm n}/I_{\rm tot} =1$.

\section{S=1 Excitations}

This part of the paper contains results for the dynamical structure factor of
 the dimerized and frustrated spin chain. At $\lambda=0$ the system consists 
of isolated dimers and therefore the total spectral weight is in the 
one-triplon channel $I_1$. Turning on $\lambda$ will reduce the spectral 
weight in the one-triplon channel and the spectral weight will also be distributed in 
the multi-triplon channels. In the unfrustrated 
chain the spectral weight is shifted almost totally from the one-triplon 
channel in the two-triplon channel on passing from strong to zero 
dimerization\cite{schmi03c}. In an analogous procedure 
we have analyzed the spectral weight distribution on the disorder line at 
$\alpha=0.5$\cite{schmi03d}. We find indications that again the two-triplon
contribution is the dominant one, even for vanishing dimerization. But
due to the complexity of the frustrated system no unambiguous extrapolations
are possible\cite{notiz2}. 
Thus we do not have a final answer for the massive frustrated phase.

In the following we will show results for the one-triplon and the two-triplon
contribution to the spectral density. The one-triplon contribution contains
most of the spectral weight at strong and intermediate dimerization. In the
limit of small dimerization it is highly reduced and becomes unimportant for
zero dimerization. We expect that the two-triplon contribution is the only 
relevant contribution in the limit of zero dimerization.

A discussion of the one-triplon contribution to the dynamic susceptibility 
of a dimerized chain without frustration can be found in a work by M\"uller
and Mikeska\cite{mulle03}. Recently Zheng et al.\  published 
results for the one- and two-triplon contribution of a strongly dimerized spin
chain without frustration \cite{zheng03a}.  Our
results at zero frustration agree with the findings of Zheng et al..

Here we want to extract the generic features of the 
two-triplon contribution for various dimerization and frustration in order to
gain insight in the evolution of this contribution in the limit of vanishing 
dimerization. Therefore it is interesting to compare our results 
with results for the dynamical structure factor for zero dimerization.

At $\alpha=0$ an exact calculation of the two-spinon contribution to the 
dynamical structure factor using Bethe-ansatz is possible\cite{karba97}. The
two-spinon contribution exhausts 72.89\% of the total spectral weight and it
displays a singular divergent behavior at the lower edge of the two-spinon continuum.
At finite frustration only numerical results using exact diagonalization at
finite temperatures including frustration are available\cite{fabri98b}. In 
addition, there are also results using abelian bosonization extracting the 
universal features of the dynamical structure factor at low energies for 
small dimerization\cite{gogol98}. In the following we will identify 
the major features of these studies in our triplonic description at finite
dimerization.
\begin{figure}[htbp]
  \begin{center}
    \includegraphics[width=\columnwidth]{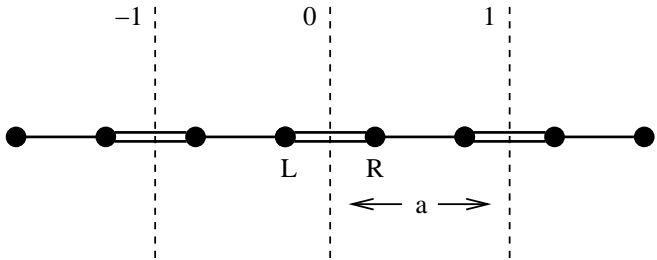} 
    \caption{Sketch of the local observable for S=1 excitations. Double lines 
denote strong bonds and single lines weak bonds. The observable couples to the
 left (L) and to the right (R) spin of a strong bond.} 
    \label{fig:Skizze_S1}
  \end{center}
\end{figure}
\subsection{One-Triplon Contribution}

The physical observable $R^{S=1}$ for total $S=1$ excitations is locally
\begin{equation}
 R^{S=1}_{\rm loc} = S^{\rm z}_i 
\end{equation}
where $i$ denotes a site of the chain. In the following we will name the left and the right site on a strong bond by $L$ and $R$, cf.\ 
Fig.~\ref{fig:Skizze_S1}.
We decompose the action of the full observable $R^{S=1}|_{\rm 1trp}$ in the one-triplon channel on the 
ground-state for fixed one-triplon momentum $k$  by writing
\begin{equation}
 R^{S=1}|_{\rm 1trp}|0\rangle = A_{\rm 1trp}^{S=1}(k)|k\rangle ,
\end{equation}
where the amplitudes $A_{\rm 1trp}^{S=1}$ are given by
\begin{eqnarray}
 A_{\rm 1trp}^{S=1} &=& \frac{1}{\sqrt{2}}\sum_l \left[a_l^{\rm L}e^{-iK(2l+\frac{1}{2})}+ a_l^{\rm R}e^{-iK(2l-\frac{1}{2})}\right]\nonumber\\
 &=& -\sqrt{2}i\sum_l a_l^L\sin \left(k(2l+\frac{1}{2})\right).
\end{eqnarray}
The sum runs over all strong bonds $l$. The coefficient 
$a_l^L$ is the amplitude for the 
creation of one triplon at site $l$ by $S_L^z$. The amplitudes of $S_R^z$ 
do not need to be calculated due to the symmetry $a_l^R=-a_{-l}^L$. The basic
unit length $a/2$ is the distance between two neighboring sites. In the figures
momenta between $0$ and $1$ are given in units of $\pi/(a/2)$. So
the comparison between our results and the conventional notation for
undimerized chains using the distance $a^\prime$ between two 
neighboring spins as unit length ($a^\prime=a/2$) is simplified.

\begin{figure*}
  \begin{center}
    \includegraphics[width=\textwidth]{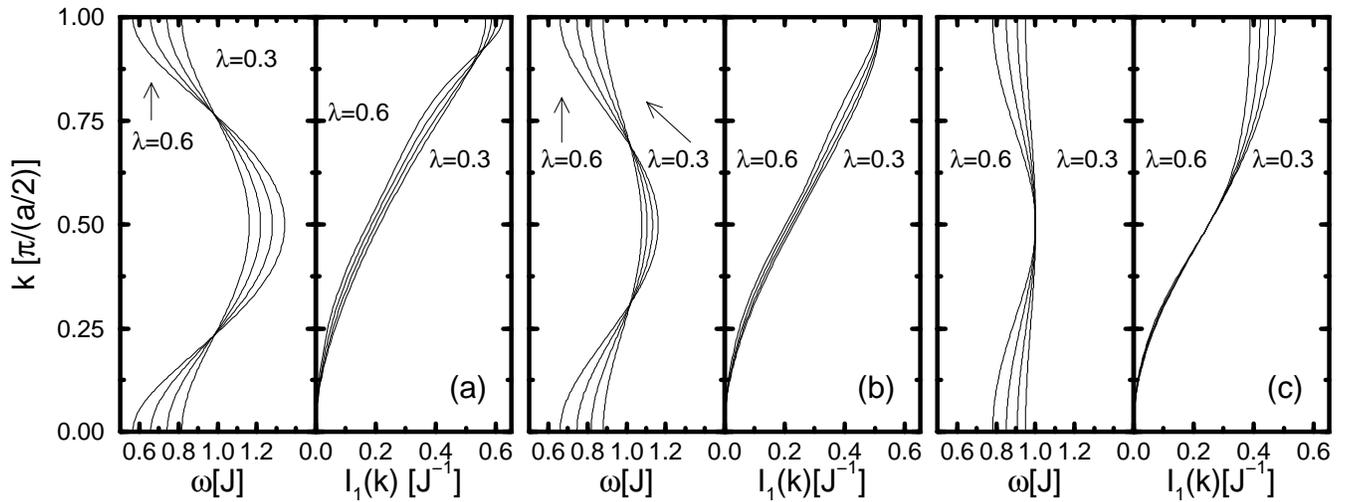}
    \caption{One-triplon dispersion $\omega (k)$ (left panels) and one-triplon spectral
 weight $I_1 (k)$ (right panels) for $\lambda=\{0.3;0.4;0.5;0.6\}$. In Fig.~(a) 
$\alpha=0.0$, in Fig.~(b) $\alpha=0.25$ and in Fig.~(c) $\alpha=0.5$.} 
    \label{fig:S1_1trp}
  \end{center}
\end{figure*}

In Fig.~\ref{fig:S1_1trp} the results for the one-triplon dispersion 
$\omega (k)$ (left panels) and the one-triplon spectral weight $I_1 (k)$ 
(right panels) are shown. We present results for $\lambda=\{0.3;0.4;0.5;0.6\}$
 and $\alpha=0$ (Fig.~\ref{fig:S1_1trp}a), $\alpha=0.25$ 
(Fig.~\ref{fig:S1_1trp}b) and $\alpha=0.5$ (Fig.~\ref{fig:S1_1trp}c).

Let us discuss first the case of vanishing frustration ($\alpha=0$). At 
$\lambda=0$ the system consists of isolated dimers and the one-triplon 
dispersion is flat. Turning on $\lambda$ the triplon starts to hop from dimer 
to dimer and it acquires a finite dispersion (Fig.~\ref{fig:S1_1trp}a, left panel).
The dispersion has minima at $k=0$ and $k=\pi$ (in units of $1/(a/2)$) 
which represent the one-triplon gap $\Delta$. In the limit of zero 
dimerization the one-triplon gap closes and it is to be expected that
the one-triplon dispersion equals the well known 
Cloizeaux-Pearson\cite{cloiz62} dispersion relation 
$\omega_{\rm CP}(k)=\pi/2|\sin (k)|$.
  
The one-triplon spectral weight $I_1 (k)$ is shown in the right panel of 
Fig.~\ref{fig:S1_1trp}a. The leading term of $I_1 (k)$ is proportional to 
$\sin^2 (k/2)$. It is called the dimer structure factor\cite{mulle03} 
(note again that we measure the momentum $k$ in units of $1/(a/2)$). The 
one-triplon spectral weight is concentrated at $k=\pi$. At finite dimerization
the reduction of $I_1$ due to the inter-dimer exchange is achieved by momenta
$k<0.9\pi$. The spectral weight increases in a small interval about 
$k=\pi$\cite{mulle03}. In the limit of smaller dimerization 
the one-triplon spectral weight becomes more concentrated about 
$k=\pi$. The total weight, integrated over momentum and frequency, vanishes
for vanishing dimerization.

In Fig.~\ref{fig:S1_1trp}b the corresponding results for $\alpha=0.25$ are 
shown. The one-triplon dispersion is similar in shape to the case of zero 
frustration. Due to the finite frustration the excitations become more local
and the triplon is less dispersive. The gap values are slightly larger and 
the maximum values of the one-triplon dispersion are slightly lower for the 
various values of $\lambda$ in comparison to the unfrustrated case 
(Fig.~\ref{fig:S1_1trp}a and Fig.~\ref{fig:S1_1trp}b, left parts).

The one-triplon spectral weight at $\alpha=0.25$ differs from the one 
at $\alpha=0$ for momenta close to $k=\pi$ (Fig.~\ref{fig:S1_1trp}b,
right part). The spectral weight is reduced for all momenta at $\alpha=0.25$ 
on increasing $\lambda$. But the reduction is smallest for $k=\pi$. In 
the limit of zero dimerization the one-triplon spectral weight $I_1 (k)$
vanishes for all momenta.
The left part of Fig.~\ref{fig:S1_1trp}c shows the one-triplon dispersion 
$\omega(k)$ for $\alpha=0.5$. The dispersion is highly reduced due to the 
increased locality of the triplon. At $k=\pi/2$, the one-triplon state is an
eigen-state of the system\cite{caspe82,caspe84} and it has an energy of $J$
independent of $\lambda$. In the limit of zero dimerization the system remains
in a gapped state\cite{shast81a,halda82b,julli83,okamo92,egger96}.

The right part of Fig.~\ref{fig:S1_1trp}c shows $I_1 (k)$ for $\alpha=0.5$.
The spectral weight is reduced for increasing $\lambda$ for all momenta 
except $k=\pi/2$. Since the one-triplon state at $k=\pi/2$  is an
eigen-state independent of $\lambda$ its  spectral weight is also
constant\cite{singh99a}. In contrast to the previous cases, there is 
also a one-triplon contribution for zero dimerization, at least for
$k=\pi/2$, but most probably also in the vicinity of this momentum.

\subsection{Two-Triplon Contribution}

In this section we discuss the two-triplon contribution to the dynamical 
structure factor. The two-triplon contribution displays additional physics in
comparison to the one-triplon part. The reason is that besides the kinetic 
part of the excitations also the triplon-triplon interaction is important and
has to be included. An attractive interaction can lead to bound states of two
triplons. Furthermore, the total momentum of two triplons does not fix the
state of the system. There is also a relative momentum between the triplons
which is not fixed. Thus there is a continuum of two-triplon states for each
given total momentum. Let us turn to the spectral properties of
the two-triplon continuum and the two-triplon bound states.
  
\begin{figure}
  \begin{center}
    \includegraphics[width=\columnwidth]{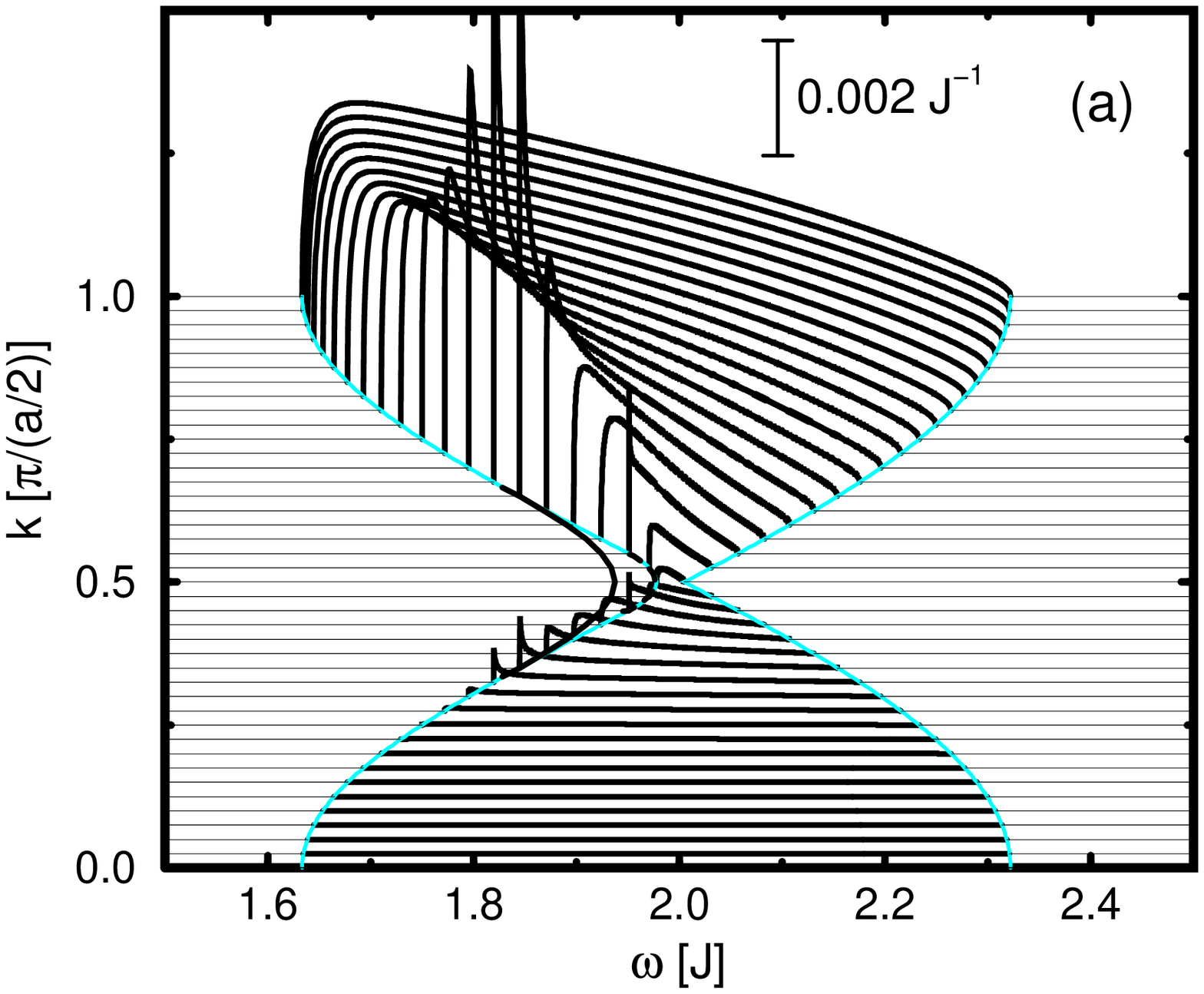}
    \includegraphics[width=\columnwidth]{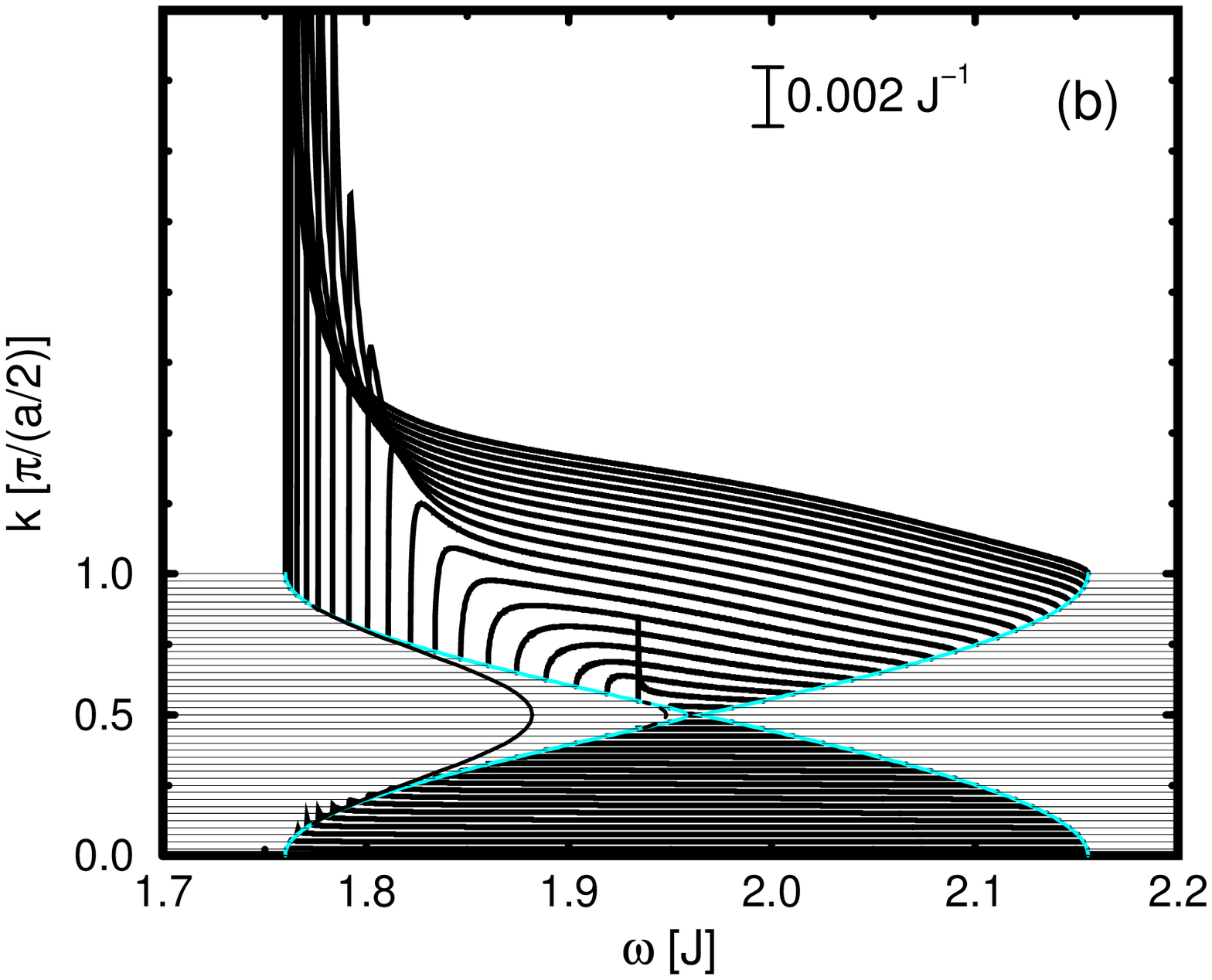}
    \includegraphics[width=\columnwidth]{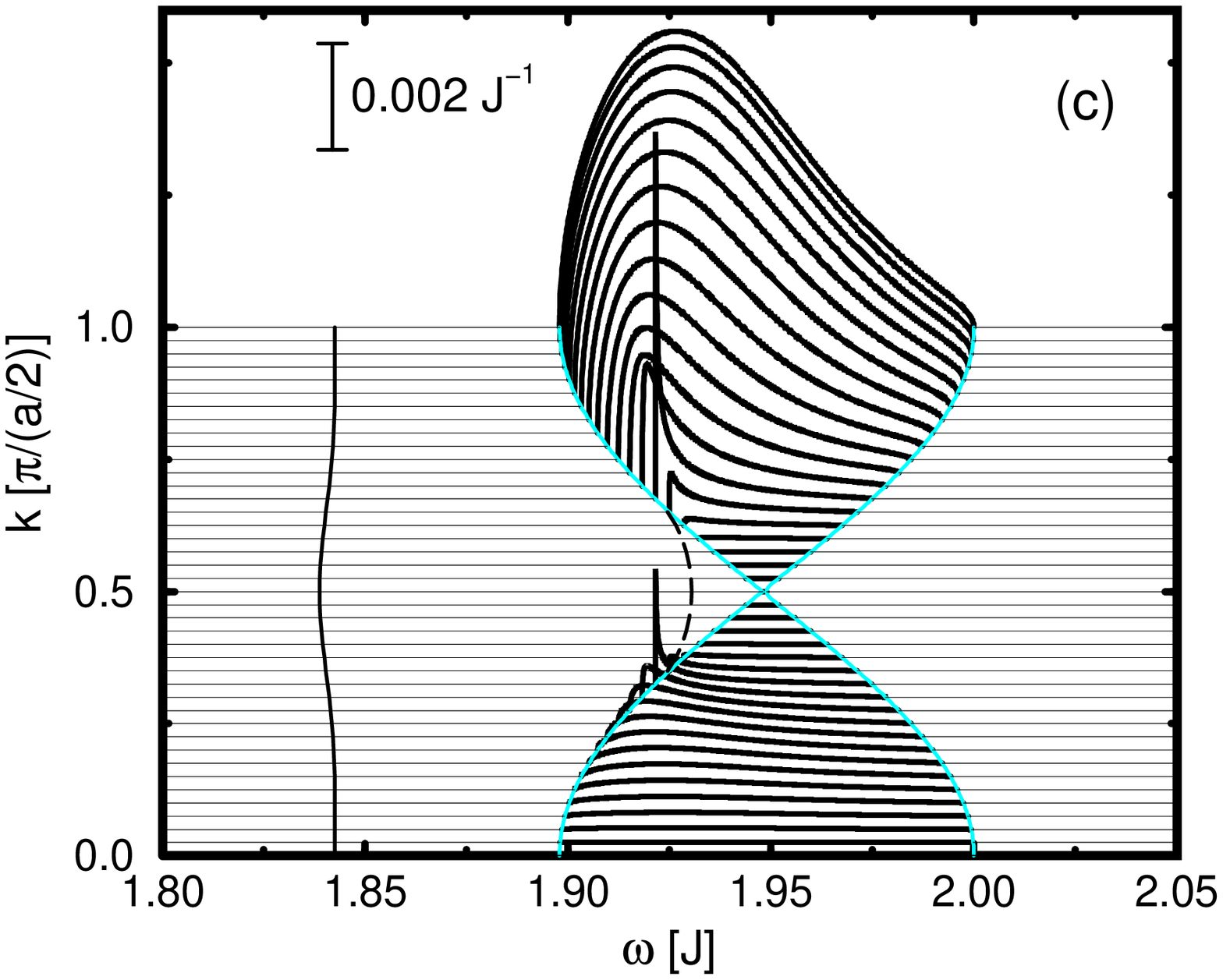} 
    \caption{Two-triplon spectral density $I_2(k,\omega)$ for $R^{S=1}$ with $\lambda=0.3$ 
    and $\alpha=0.0$ (a), $\alpha=0.25$ (b) and $\alpha=0.5$ (c). 
    Gray lines denote the lower
    and upper bound of the continuum. Black lines indicate dispersion of 
    two-triplon bound states.} 
    \label{fig:Spectral_D_S1_l03}
  \end{center}
\end{figure}
\begin{figure*}
    \includegraphics[width=\textwidth]{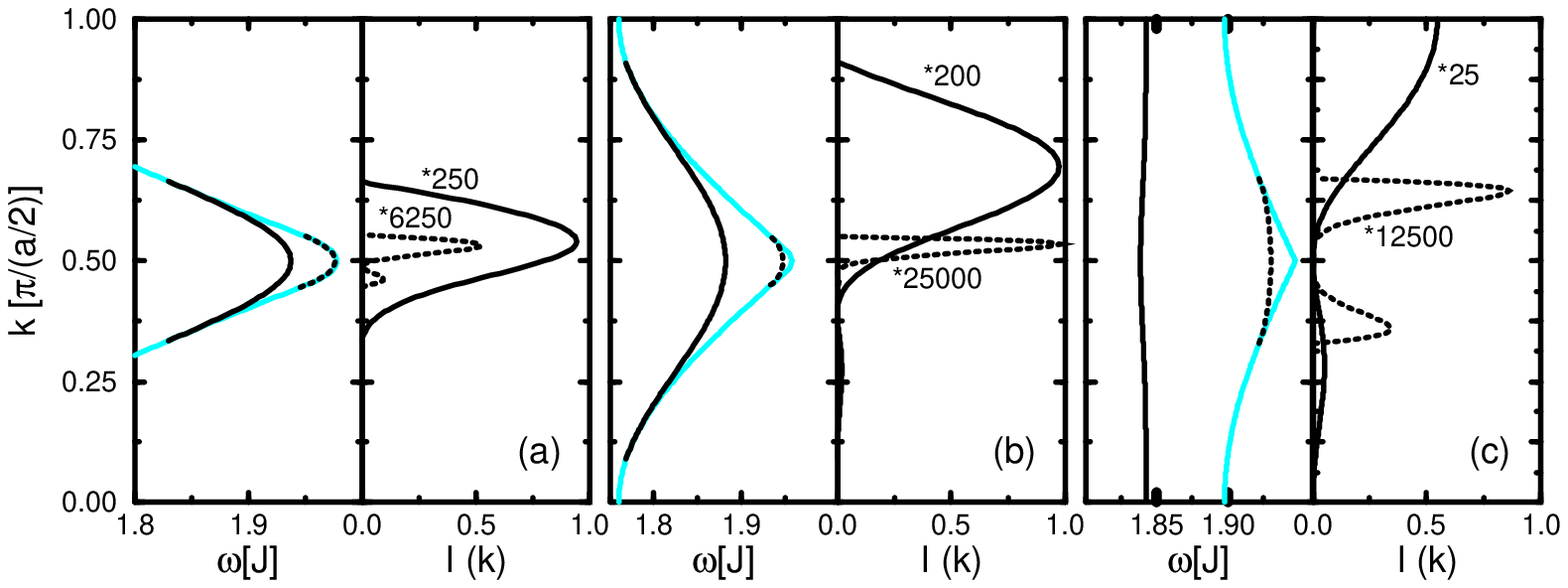}
    \caption{Two-triplon bound states for $R^{S=1}$ with $\lambda=0.3$ and $\alpha=0.0$ 
(a), $\alpha=0.25$ (b) and $\alpha=0.5$ (c). Left panels show the dispersion 
of the bound states; right panels show the spectral weights of the bound
 states multiplied by the indicated factors. Gray lines denote lower
bound of the continuum.} 
    \label{fig:BS_D_S1_l03}
\end{figure*}
\begin{figure}
  \begin{center}
    \includegraphics[width=\columnwidth]{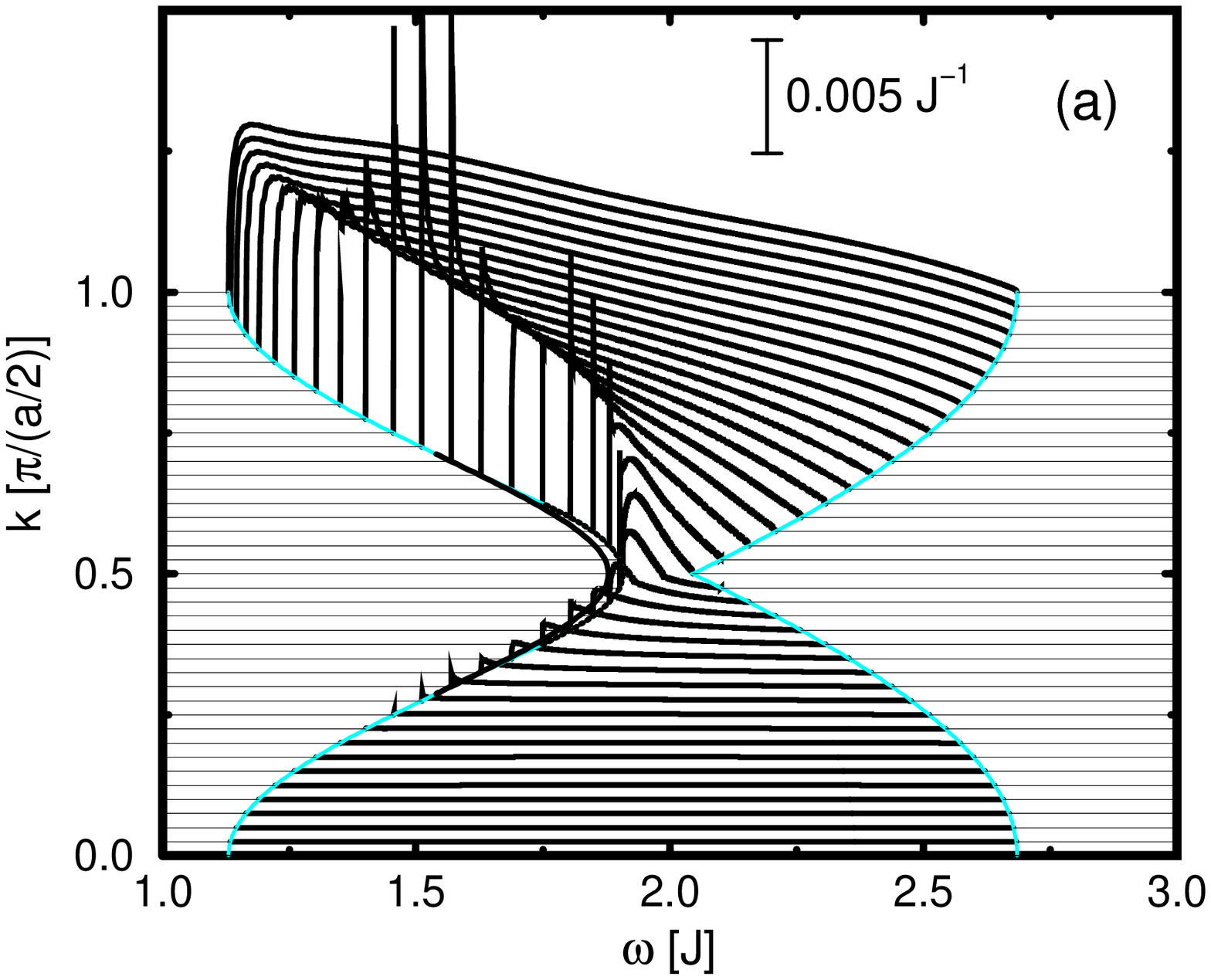} 
    \includegraphics[width=\columnwidth]{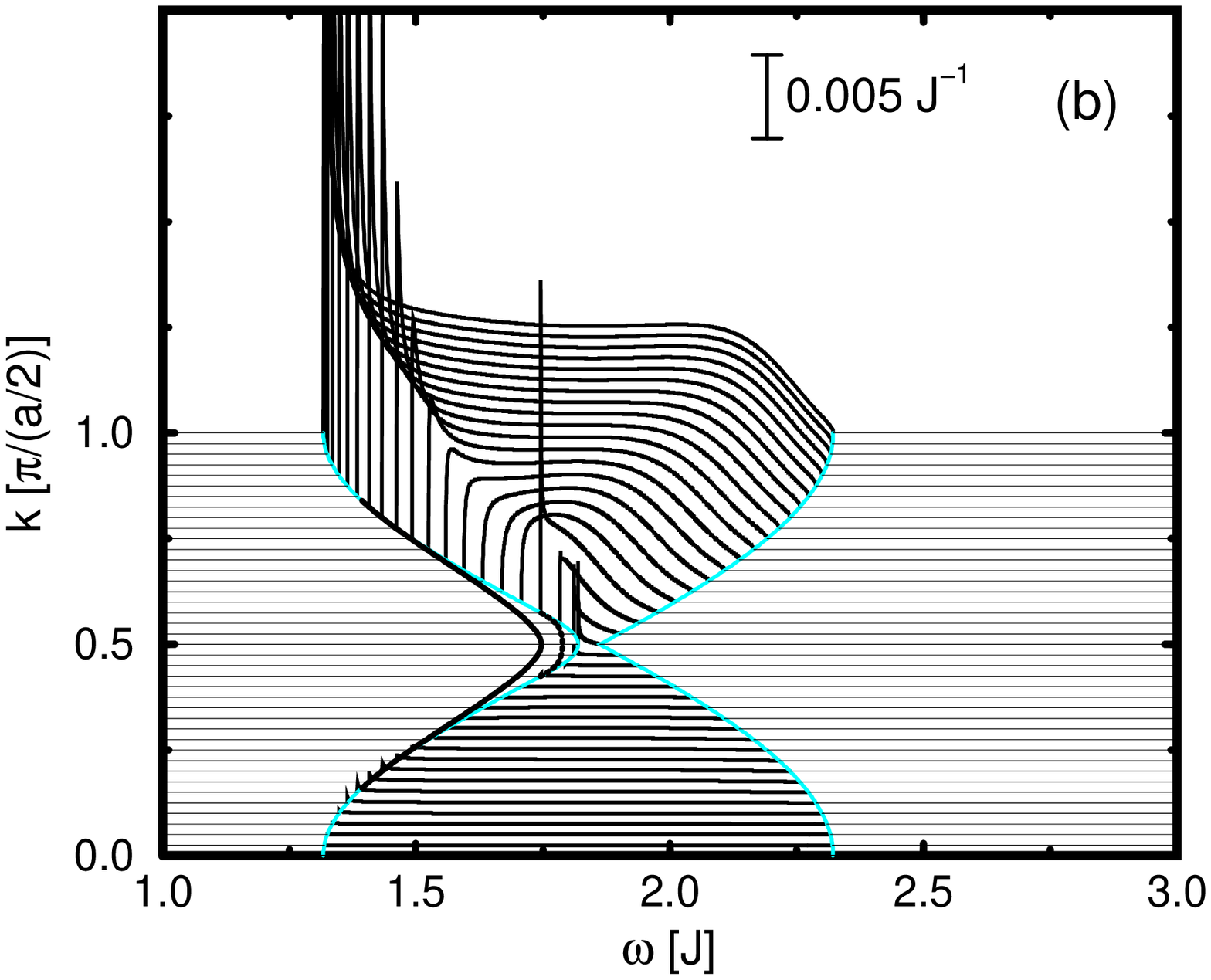}
     \includegraphics[width=\columnwidth]{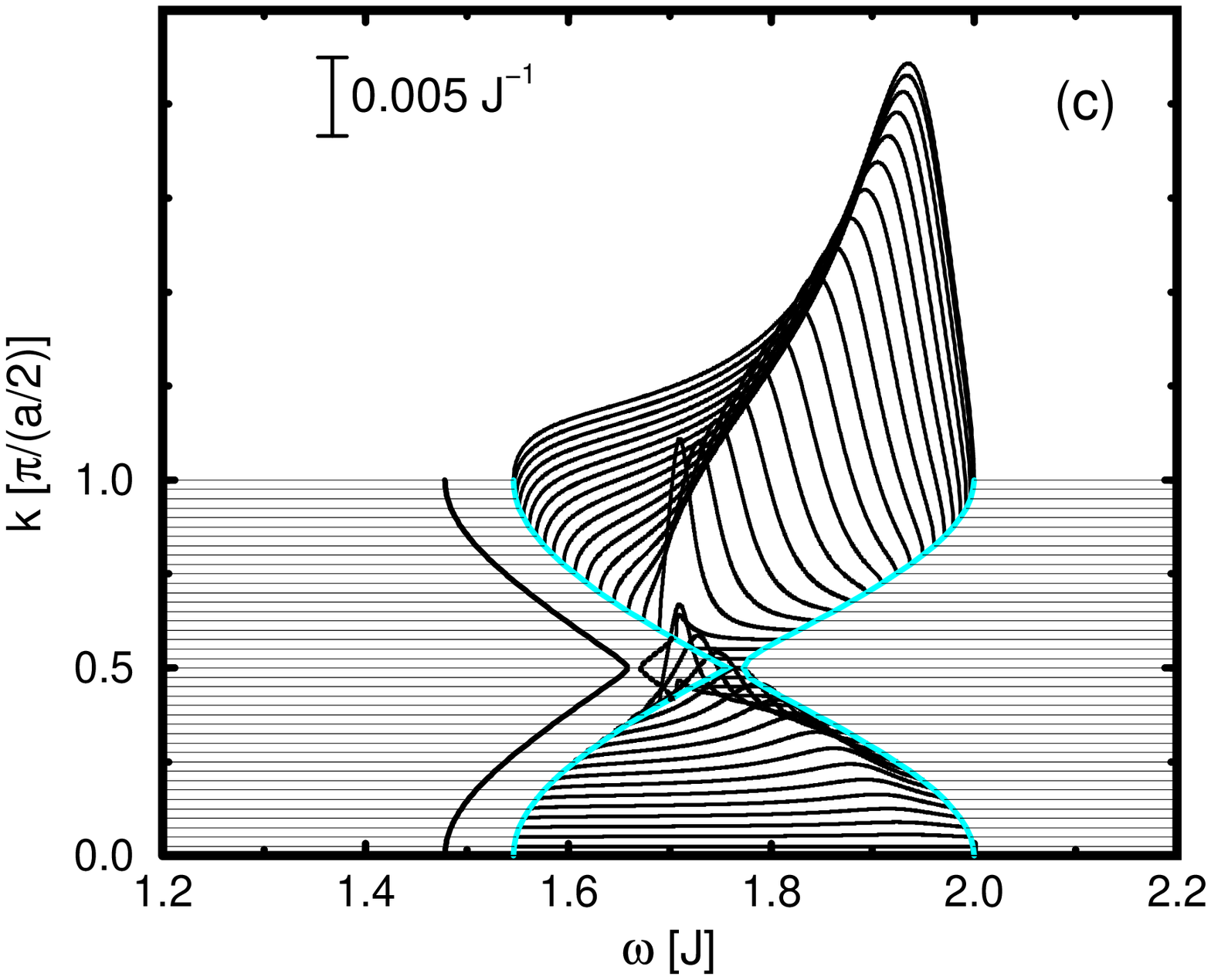} 
    \caption{Two-triplon spectral density $I_2(k,\omega)$ for $R^{S=1}$ with $\lambda=0.6$ 
and $\alpha=0.0$ (a), $\alpha=0.25$ (b) and $\alpha=0.5$ (c). 
    Gray lines denote lower
 and upper bound of the continuum. Black lines indicate dispersion of 
two-triplon bound states.} 
    \label{fig:Spectral_D_S1_l06}
  \end{center}
\end{figure}
\begin{figure*}
    \includegraphics[width=\textwidth]{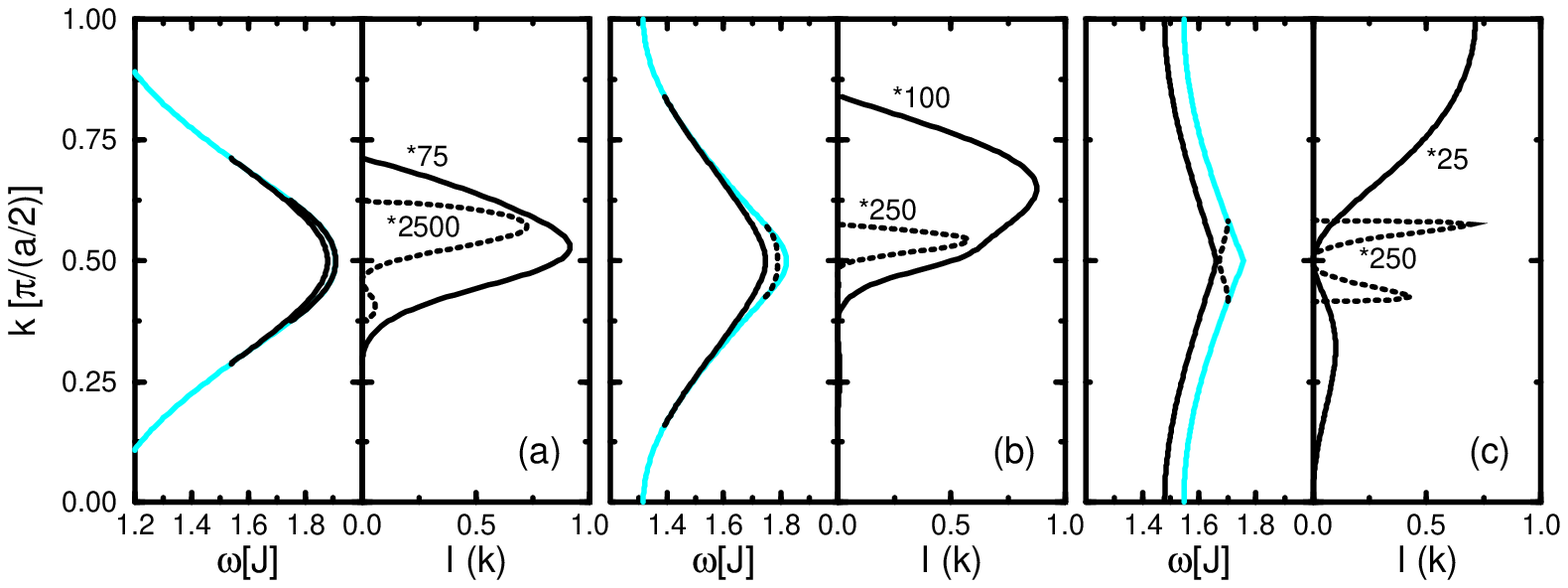}
    \caption{Two-triplon bound states for $R^{S=1}$ with $\lambda=0.6$ and $\alpha=0.0$ 
(a), $\alpha=0.25$ (b) and $\alpha=0.5$ (c). Left panels show the dispersion
 of the bound states; right panels the spectral weights of the bound
 states multiplied by the indicated factors. Gray lines denote lower
bound of the continuum.} 
    \label{fig:BS_D_S1_l06}
\end{figure*}

We decompose the action of the full observable $R^{S=1}|_{\rm 2trp}$ in the two-triplon channel on the ground-state
 for fixed two-triplon momentum $k$ in the two-triplon sector
\begin{equation}
 R^{S=1}|_{\rm 2trp}|0\rangle = \sum_d A_{\rm 2trp}^{S=1}(k,d)|k,d\rangle .
\end{equation}
Here $d$ denotes the relative distance between the two triplons and 
\begin{equation}
 A_{\rm 2trp}^{S=1}(k,d)=-\sqrt{2}i\sum_l a_{l,l+d}^{L,S=1}\sin \left(k(2l+\frac{1}{2}+d)\right) .
\end{equation}
The sum runs over all strong bonds $l$ and $a_{l,l+d}^{L,S=1}$ is the 
amplitude for the creation of two triplons on dimers $l$ and $l+d$ by 
$S^{\rm z}_{L}$. 
Here it is convenient to use a mixed representation
in which the center-of-mass coordinate is Fourier transformed and the
relative coordinate is dealt with in real space.
The action of $S^{\rm z}_{R}$ does not need to be 
calculated due to the relation $a_{l,l+d}^{R,S=1}=-a_{-l-d,-l}^{L,S=1}$. 
The basic unit length $a/2$ is again the distance between two neighboring 
sites. The spectra show results for momenta between $0$ and $1$ in 
units of $\pi/(a/2)$ (see Fig.\ref{fig:Skizze_S1}).
In Figs. \ref{fig:Spectral_D_S1_l03}-\ref{fig:BS_D_S1_l06} the results for the
two-triplon continua, the  dispersions of the two-triplon bound states and 
their  spectral weights are shown.

Fig.~\ref{fig:Spectral_D_S1_l03} displays the spectral density of the 
two-triplon continuum as a function of frequency and momentum. The spectrum for 
fixed momentum $k$ is shifted by $k$ in y-direction in order to provide a 
three dimensional view on the spectral densities. The lower and upper band 
edge is marked by solid gray lines. If there are any two-triplon bound states,
their dispersion is also displayed as a black line. We denote the $S=1$
two-triplon bound states as $T_n$ and the $S=0$ two-triplon bound states 
as $S_n$ where $n=\{1,2,\ldots\}$.

Detailed information about the bound states is given in 
Fig.~\ref{fig:BS_D_S1_l03} which consists of two parts for each parameter set.
The left part shows an enlargement of the dispersion of the bound state 
and the lower bound of the two-triplon continuum. 
In the right part the corresponding spectral 
weight of the bound states is shown. The spectral weight is multiplied by the
given factors for clarity.

What are the general features of the obtained spectra? Due
to the conservation of the total $S^z$-component 
there is no spectral weight at zero momentum. 
The energies of the system possess a reflection symmetry about $k=\pi/2$ which
is a consequence of the inversion symmetry $k \leftrightarrow -k$ and of the 
coupling of the momenta $k$ and $k+\pi$\cite{uhrig96b}. This symmetry can be 
clearly seen in the bound state energies and in the lower and the upper band 
edges of the continuum. It does {\em not} hold for the spectral
weights\cite{uhrig96b}.

In Fig.~\ref{fig:Spectral_D_S1_l03}a the spectral density for $\lambda=0.3$ 
and $\alpha=0.0$ is shown. The spectral weight is mostly concentrated at the 
lower band edge of the continuum. There are two bound states centered about 
$k=\pi/2$ leaving the continuum at some finite momentum. The dispersions and 
the spectral weights of the bound states are plotted in 
Fig.~\ref{fig:BS_D_S1_l03}a. The points where the bound states are leaving the
continuum can also be discerned in the singular behavior of the lower band 
edge of the continuum. 
The spectral weight is mainly concentrated in the first bound 
state $T_1$. The spectral weight of the second bound state $T_2$ is 
highly reduced.

The binding energy of the bound states has its maximum
at $k=\pi/2$. It vanishes quadratically $\propto (k-k_c)^2$ when the bound state enters
the continuum. Correspondingly their spectral weight vanishes linearly $\propto |k-k_c|$
in accordance with the exemplary calculation provided in Ref.~\cite{schmi03c}.
There it was shown for square root type continua
that the binding energy vanishes quadratically as function of
an external parameter while the spectral weight of the bound state
vanishes linearly. The external parameter was the attraction strength;
in the present case it is the total momentum which controls the relative strength
of interaction and kinetic energy.

Decreasing the dimerization to intermediate values ($\lambda=0.6$, see 
Figs.~\ref{fig:Spectral_D_S1_l06}a and \ref{fig:BS_D_S1_l06}a), there are no
qualitative changes in the spectrum. The spectral weight is mainly 
concentrated at low energies. The range of energies, where the bound states
exist, is slightly enhanced while the binding energy is slightly reduced.
We expect that this tendency continues to lower values of the dimerization.
For exactly zero dimerization, but {\em not} for an arbitrarily small but 
finite one, the first bound state $T_1$ 
coincides with the lower bound of the continuum leading to a square root 
divergence at the lower edge of the continuum for all momenta. Our  
expectation is strongly corroborated by the exact results for the 
spectral densities in the sine-Gordon model\cite{gogol98,smirn92}. 
The generic behavior is a square root behavior at the band edges. Only
if a breather becomes degenerate with the multi-particle band edge
the square root behavior switches to a square root divergence.
Exactly the same characteristics appears naturally in the triplonic description.
Recall also that for the uniform spin chain at zero dimerization the 
square root divergence is well-known from the exact two-spinon contribution 
to the dynamical structure factor\cite{karba97}.

In Figs.~\ref{fig:Spectral_D_S1_l03}b and \ref{fig:Spectral_D_S1_l03}c the 
spectra for strong dimerization and finite frustration $\alpha=0.25$ and 
$\alpha=0.5$ are shown. The corresponding information about the two-triplon
bound states is plotted in Figs.~\ref{fig:BS_D_S1_l03}b and 
\ref{fig:BS_D_S1_l03}c. The frustration makes the
excitations more local and less dispersive 
which leads to a narrowing of the two-triplon continuum. 
At the same time the triplon-triplon interaction is enhanced causing 
an increase of  the binding energy of the bound states. It can be nicely 
seen that for $\alpha=0.25$ the first bound state extends over a wide range 
in momentum space lying for small and large momenta very close to
the lower band edge inside the 
continuum. For $\alpha=0.5$ the bound state $T_1$ exists for all momenta.

Due to the existence of the bound states only in a finite interval of momentum
for both values $\alpha=0$ and $\alpha=0.25$ 
the qualitative distribution of their spectral weight is similar. This will be
true for all values of frustration between $0$ and $0.25$.
The same holds for the bound state $T_2$ for $\alpha=0.5$. In contrast, the 
spectral weight of the first bound state $T_1$, which is well separated from 
the continuum for $\alpha=0.5$, has its maximum at $k=\pi$.

The whole two-triplon contribution, i.e. the two-triplon bound 
states and the two-triplon continuum, vanishes for $k=\pi/2$ at $\alpha=0.5$.
Here the one-triplon excitation is an exact eigen-state of the spectrum and 
therefore comprises the total spectral weight\cite{caspe82,caspe84} (see also preceding Section).

We now turn to the influence of the frustration on the shape of the 
two-triplon continuum. In the case of vanishing frustration the spectral 
weight is distributed mainly close to the lower band edge for strong and 
intermediate dimerization. The spectral weight decreases monotonically for
higher energies. Turning on the frustration we observe a shift of spectral 
weight towards higher energies. In the case of strong dimerization this 
tendency is weak (Figs.~\ref{fig:Spectral_D_S1_l03}b and 
\ref{fig:Spectral_D_S1_l03}c) while for intermediate dimerization we observe 
a huge transfer of spectral weight (Figs.~\ref{fig:Spectral_D_S1_l06}b and 
\ref{fig:Spectral_D_S1_l06}c). This transfer produces a non-monotonic shape 
for intermediate dimerization and $\alpha=0.25$ having a minimum of spectral
 weight inside the continuum (Fig.~\ref{fig:Spectral_D_S1_l06}b). Increasing 
the frustration ($\alpha=0.5$) shifts the minimum to the lower band edge. The
 spectral weight is mainly at the upper band edge 
 (Fig.~\ref{fig:Spectral_D_S1_l06}c).

These observations are very similar to the results obtained by exact 
diagonalization at finite temperatures for the dynamical structure factor of
 a homogeneous spin chain including frustration\cite{fabri98b}. There a 
decrease of spectral weight at $k=\pi$ inside the continuum is observed on 
increasing the frustration. This results in a high-energy maximum for large 
frustration and a minimum inside the continuum, i.e., a trough-like shape.
From this comparison we conclude that our findings represent the generic 
features which are also valid in the limit of vanishing dimerization.

\subsection{Comparison with Field Theory Results}

In the first part of this section we concentrate on the $S=1$
breather and the corresponding behavior of the lower band edge of the
two-triplon continuum. In the second part we have a closer look
at the energy of the $S=0$ breather and at the one-triplon gap.
Finally, the importance of marginal terms for the quantitative
shape of spectral densities is discussed. A detailed analysis of the $S=0$ 
channel which contains also a discussion of the singlet two-triplon bound 
states is presented in the next section. 
     
Let us first look at the results obtained from bosonization and the continuum
 limit renormalization group approach\cite{affle86b,affle87}. 
The bosonized form of the dimerized and frustrated spin chain reads
\begin{eqnarray}
 \label{H_FT}
 H_{\rm FT}&=&\frac{v}{2\pi}\int_{-\infty}^{\infty}\left[ 
K\left(\pi\Pi\right)^2+K^{-1}\left(\partial_x\Phi\right)^2\right]dx 
\nonumber \\
&& +\int_{-\infty}^{\infty}\left[\delta A\cos(2\Phi)+D\cos(4\Phi) \right]dx ,
\end{eqnarray}
where $v$ is the spin-wave velocity and $K=1/2$ the interaction parameter for 
the isotropic chain. The value of the critical frustration 
$\alpha_{0,c}=0.241167$ depends on the physics at short distances and is only
accessible by numerical techniques\cite{julli83,okamo92,casti95,egger96}. 
The $\delta\cos(2\Phi)$ term is strongly relevant while the
$D\cos(4\Phi)$ term is marginally irrelevant for $\alpha_0<\alpha_{0,c}$ and 
marginally relevant for $\alpha_0>\alpha_{0,c}$. It is commonly
accepted and numerically confirmed\cite{chitr95,white96,chitr97,bouze98a} 
that the marginal term can be neglected best for $\alpha_0=\alpha_{0,c}$
(for further discussion see below).
Then a simple sine-Gordon model at $K=1/2$ remains of which the
spectral densities are known \cite{gogol98,smirn92}.

The $S=1$ response function displays a square root divergence
$\propto (\omega-\omega_0)^{-1/2}$
at the lower band edge. Here the energy of the $S=1$ breather 
$\omega_{{\rm br},S=1}$ is exactly degenerate with the lower band edge 
($\omega_{{\rm br},S=1}=2\Delta$). This is in agreement with what we find
at $\alpha=0.25$, cf.~Figs.~\ref{fig:Spectral_D_S1_l03}b and 
\ref{fig:Spectral_D_S1_l06}b. 
Without any frustration, however, we find
a square root behavior $\propto (\omega-\omega_0)^{1/2}$
, cf.~Figs.~\ref{fig:Spectral_D_S1_l03}a and \ref{fig:Spectral_D_S1_l06}a.
Hence we conclude that the sine-Gordon model does not describe
the unfrustrated, dimerized spin chain exhaustively. But the sine-Gordon model
applies to the spin chain at critical frustration where
the higher cosine-term $\cos(4\Phi)$ vanishes. It is interesting to note that 
we find a square root divergence for $\alpha\approx\alpha_{0,c}$ and not
for $\alpha_0\approx\alpha_{0,c}$. 

The conclusion about the applicability of the sine-Gordon model to the 
unfrustrated and dimerized spin chain is in agreement with the results of
the numerical investigation of the bound states \cite{bouze98a}. It is known 
that the elementary excitations of the SU(2) symmetric 
sine-Gordon model consist of soliton and antisoliton excitations and two breathers, 
bound states, plus one breather which is degenerate with the
lower band edge\cite{dashe75b,halda82b,uhrig96b}. 
The lowest-lying breather is degenerate with the soliton 
and antisoliton excitations and corresponds to the $S^z=0$ triplet state
in spin language. This fixes the interaction parameter to $K=1/2$. The second
breather is assigned to a spin singlet excitation, since there is no counter 
part in the soliton or antisoliton sector. The ratio between the 
energy of the $S=0$ breather $\omega_{{\rm br},S=0}$ and the one-triplon gap 
$\Delta$ is exactly $\sqrt{3}$ at $K=1/2$.

\begin{figure}[htbp]
  \begin{center}
    \includegraphics[width=\columnwidth]{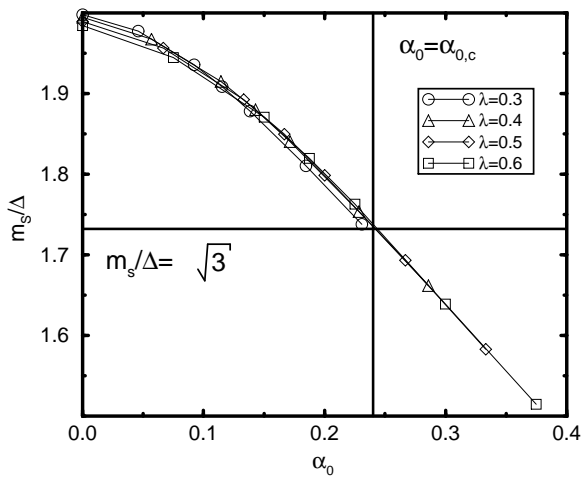} 
    \caption{Ratio of the singlet two-triplon bound state energy $m_S$ over 
the one-triplon gap $\Delta$ for $\lambda=\{0.3;0.4;0.5;0.6\}$ depending on 
the frustration $\alpha_0$. Horizontal black solid line denotes $\sqrt{3}$ and
 vertical black solid line shows the critical frustration $\alpha_{0,c}$.} 
    \label{fig:Breather_S0}
  \end{center}
\end{figure}
 In Fig.~\ref{fig:Breather_S0} this ratio is shown for various values of 
$\lambda=\{0.3;0.4;0.5;0.6\}$ versus the bare frustration $\alpha_0$. We have
used Pad\'{e} extrapolants for $\lambda=0.6$. It can be clearly seen that 
almost all points (except the case for strong dimerization $\lambda=0.3$) fall
onto one curve. The values for $\alpha_{0,c}$ and $\sqrt{3}$ are included as 
solid lines. The point where these two lines cross lies on
the calculated curve and corresponds to the prediction of the 
SU(2) symmetric sine-Gordon model. In all other 
cases $(\alpha_0\neq\alpha_{0,c})$ the ratio $\omega_{{\rm br},S=0}/\Delta$ 
differs from $\sqrt{3}$. This is due to 
corrections resulting from the marginal term $\cos(4\Phi)$. Our calculations 
agree perfectly with previous numerical results\cite{bouze98a}.
       
The relative importance of the two cosine-terms in Eq.~(\ref{H_FT})
are at finite dimerization is a subtle issue which we discuss in the following.

In the self-consistent harmonic approximation 
\cite{nakan80,nakan81} we replace
$\Phi \to \Phi_{\rm class} + \hat\Phi_{\rm fluct}$ where only
the fluctuation part is of operator character. The cosine-terms
can then be approximated by
\begin{eqnarray}
&&\cos(n \Phi) \quad \to \\ \nonumber
&&\exp(-(n^2/2) \sigma(x)) 
\cos(n\Phi_{\rm class}(x))
       \left(  1-\frac{n^2}{2}\hat\Phi_{\rm fluct}^2 \right)\ .
\end{eqnarray}
where $\sigma (x):= \langle \hat\Phi_{\rm fluct}(x)^2\rangle$. This kind
of approach corresponds to renormalization in first order.
In the ground state (without solitons) one has $\Phi_{\rm class}=0$.
In the ungapped phase the fluctuations diverge, hence $\sigma\to \infty$.
But in the gapped phase, the fluctuations are cutoff at low energies
so that $\sigma = -(K/2)\ln (\Delta/\Delta_0)$ where $\Delta$ is the gap
and $\Delta_0$ is proportional to  the ultraviolet cutoff. 

Since the square of the gap $\Delta^2$ is proportional to the coefficient
of $\hat\Phi(x)^2$ one obtains from the term
$\delta \cos(2\Phi)$ the self-consistency equation 
\begin{eqnarray}
\nonumber
\Delta^2 &\propto& \delta\exp(-2\sigma)\\
\Delta &\propto& \delta^ {1/(2-K)}
\end{eqnarray}
yielding the well-known $\Delta \propto \delta^{2/3}$ by Cross and Fisher
\cite{cross79}.
Hence  the total contribution of this cosine-term is proportional to 
$\Delta^2$ or $\delta^{4/3}$.
The crucial point to note is that the amplitude of the second cosine-term
$\cos(4\Phi)$ is of the same magnitude $\exp(-8\sigma)\propto \Delta^{4K}$
which also yields $\Delta^2$ or $\delta^{4/3}$ for $K=1/2$ \cite{notiz4}.
Hence the self-consistent harmonic approximation tells us that even 
in the regime where the frustration is marginally irrelevant
$\alpha < \alpha_c$ it influences the low-energy physics on the quantitative
level. 

Considering, however, the renormalization to second order the marginally term is reduced logarithmically, hence the name ``marginally irrelevant'' (Refs. 34,64). The flow is valid down to the infrared cutoff which is in our case the energy gap due to the dimerization. Note that we are working in the thermodynamic limit $L=\infty$. Hence the suppression of the marginally irrelevant cosine term to the relevant one is logarithmically in $\delta$ and so very slow. Thus it is possible that the scales which we discuss in this paper ($\delta$ down to 0.25) are still too large to see the emergence of the physics of a pure sine-Gordon model with one cosine only. But we find it puzzling that no precursors of the convergence to the predictions of the pure sine-Gordon model can be seen. The ratio of the energy gaps shown in Fig. \ref{fig:Breather_S0} (or those obtained in Ref. 37) appears almost idependent of $\delta$ as far as we could investigate.     

Note that the vanishing of the $\cos(4\Phi)$ term at the end of the flow of $l$ is {\em not} identical to the question whether this term is present at the beginning of the flow. We presume that this difference explains the puzzling finding that the bare coefficient of the Umklapp term $\cos(4\Phi)$ vanishes\cite{halda82b} at $\alpha_0=1/6$, that means relatively far from the quantum critical point $\alpha_0=0.241$.

Our results in 
Figs.~\ref{fig:Spectral_D_S1_l03}-\ref{fig:Spectral_D_S1_l06}(a-b) show 
that the square root divergence known from the sine-Gordon model 
\cite{gogol98,smirn92} is changed to normal square root behavior.
So the quantitative changes of the low-energy
Hamiltonian influence the shape of the spectral densities qualitatively.
Hence for spectral densities one must know whether the effective low-energy
is a (simple) sine-Gordon model or a double sine-Gordon model with
two cosine terms. The answer depends on the energy scale considered, i.e. the value of $\delta$. 

For strong frustration $\alpha=0.5$, cf.~Figs.~\ref{fig:Spectral_D_S1_l03}c 
and \ref{fig:Spectral_D_S1_l06}c,
the physics is dominated by bound states. Their number proliferates
for decreasing dimerization \cite{notiz3,notiz4}. At $\delta=0$ there are 
infinitely many bound states densely distributed between $\Delta$
and $2\Delta$. They form the continuum which can be understood as
two-spinon continuum \cite{shast81a,notiz3,notiz4}.
The values of $\lambda$ for which we
display the spectral densities in Figs.~\ref{fig:Spectral_D_S1_l03}c and 
\ref{fig:Spectral_D_S1_l06}c are still too
low, i.e., too far in the dimerized regime to see more than two bound
states. This was also observed by Zheng {\it et al}.\cite{zheng01b}. They found in a series expansion up to order $\lambda^{19}$ three singlet and three triplet bound states.  

We attribute the fact, that only a small number of bound states
could be found so far, to the limited range of the effective interaction.
In the perturbative approaches for $\alpha=0.5$  an order of
$\lambda^n$ corresponds to a maximum range of $[n/2]$. So even calculations
at $n=19$ provide only a potential of small finite range which does
not allow for many bound states. The alternative presumption\cite{zheng01b}
that the lacking bound states is found in the channels with more than two
triplons would require that the spectral weight is passed to channels with
more and more triplons. No channel with only a finite number of bound
states may retain a finite spectral weight at zero dimerization since
at zero dimerization only a continuum is found\cite{shast81a}. 
None of our results is in favor of this scenario so that we are convinced
that the range of the interaction is the crucial point. The fact that an expansion to higher order finds more bound states in the two-triplon sector supports the view that the range of the interaction matters. But the precise
description of the deconfinement transition for vanishing dimerization
is still an open issue. Future developments like self-similar
realizations of the continuous unitary transformations will help
to release this constraint on the range of the interaction\cite{knett03a,reisc03a}.

Concerning the present data at strong frustration, 
an important trend is that on increasing $\lambda$ the spectral
weight is shifted towards higher energies.

\section{S=0 Excitations}

In this section we concentrate on the two-triplon contribution to the spectral
density with total spin zero which is relevant for optical experiments. For 
$S=0$ one needs at least two triplons which form together a state with 
vanishing total spin. For the case of isolated dimers ($\lambda=0$) the total
spectral weight is in the two-triplon contribution. At finite $\lambda$ the 
spectral weight in the two-triplon channel is reduced and spectral weight
is also found in channels with more than two triplons.

In the limit of vanishing dimerization for zero frustration we can show in a 
similar analysis as for the dynamical structure factor that the two-triplon 
contribution possesses almost the total spectral weight
\cite{schmi03c,schmi03d}. The two-triplon contribution is therefore the only 
sizable contribution for the whole range of dimerizations for the unfrustrated
case. In presence of frustration the analogous analysis is quantitatively
more difficult as stated before. But there are again
indications that only a small number of triplons dominates the spectral 
properties. Therefore, we investigate the leading two-triplon contribution for the $S=0$ channel.
For the dimerizations treated in this work there is no doubt that 
the two-triplon contribution is the only sizable term. The crucial point,
however, is to which extent we can obtain the generic features which 
govern also the limit of vanishing dimerization. All results obtained
so far show that the one- and two-triplon contributions capture indeed 
the relevant physics.
   
\begin{figure}[htbp]
  \begin{center}
    \includegraphics[width=\columnwidth]{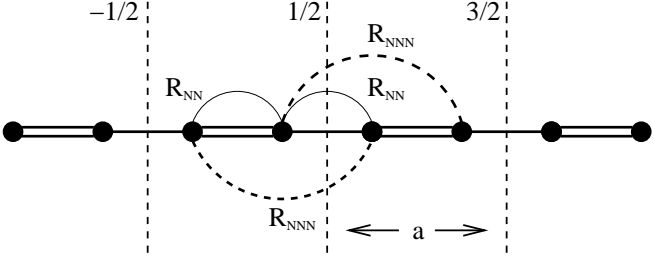} 
    \caption{Sketch of the local observables for S=0 excitations. 
      $R^{S=0}_{\rm NN}$ is a sum of couplings on strong bonds (double lines)
      and weak bonds (single line). $R^{S=0}_{\rm NNN}$ couples next nearest 
      neighbor spins.} 
    \label{fig:Skizze_S0}
  \end{center}
\end{figure}

The physical observable $R^{S=0}$ for $S=0$ excitations  locally reads
\begin{eqnarray}
 R^{S=0}_{\rm loc}& =& R^{S=0}_{\rm loc,NN}+\beta R^{S=0}_{\rm loc,NNN} ,
\end{eqnarray}
i.e., it is a sum over nearest neighbor (NN) and next nearest neighbor 
coupling (NNN). The coefficient $\beta$ is a measure for the relative 
different strength between the two couplings. It depends on the underlying 
microscopic physics and will not be discussed in this work. As  illustrated
in Fig.~\ref{fig:Skizze_S0} these observables are given by
\begin{eqnarray}
   R^{S=0}_{\rm loc,NN} & =& (1+\gamma){\bf S_{\rm 0,L}S_{\rm 0,R}}+
(1-\gamma){\bf S_{\rm 0,R}S_{\rm 1,L}} 
\end{eqnarray}
for nearest neighbor (NN) coupling and
\begin{eqnarray}
   R^{S=0}_{\rm loc,NNN}& =& {\bf S_{\rm 0,L}S_{\rm 1,L}}+{\bf S_{\rm 0,R}
S_{\rm 1,R} } 
\end{eqnarray} 
for next-nearest neighbor (NNN) coupling where $\gamma$ is proportional to 
the dimerization $\delta$. $R^{S=0}_{\rm loc,NN}$ is a sum over couplings on
 weak and strong bonds. 

We will restrict our discussion to the case of nearest neighbor coupling on 
the weak bonds and on the next-nearest neighbor coupling.
This choice is motivated by the relevance of various observables
for Raman spectroscopy and infrared absorption in the limit of 
vanishing dimerization. Raman spectroscopy  measures excitations  with 
total momentum zero while infrared absorption is governed by the response at 
large momenta\cite{loren95a,loren95b}.
 
In the case of a uniform spin chain without frustration 
the nearest neighbor Raman operator commutes with the Hamiltonian and one 
obtains a vanishing Raman response. Therefore the next-nearest neighbor Raman
 operator is the leading contribution in terms of a Loudon-Fleury scattering 
theory\cite{fleur68,shast90b}. In contrast $R_{\rm NN}$ does not commute for finite 
momenta with the Hamiltonian and will be the most important contribution to 
the infrared absorption. For simplicity, we do not 
treat $R_{\rm NN}$ completely but only the weak-bond part. 
This is no major restriction because we are interested in the generic 
properties of these quantities. In addition, the weak-bond part dominates 
for strong dimerization. 

We will discuss these two observables $R_{\rm NN}$ and $R_{\rm NNN}$
separately. For a direct comparison with experimental data one 
should take the sum over all contributing parts 
of $R^{S=0}(k)$ to account for possible interference effects.
The necessary superposition, however, depends strongly on the details of
the system and cannot be discussed generally.
 
The action of the full observable on the ground-state is decomposed
again for fixed total momentum $k$ in the two-triplon sector by
\begin{subequations}
\begin{eqnarray}
 R_{\rm NN,weak}^{S=0}(k)|0\rangle &=& 
\sum_d A_{\rm 2trp,NN,weak}^{S=0}(k,d)|k,d\rangle\quad  \\
 R_{\rm NNN}^{S=0}(k)|0\rangle &=& 
\sum_d A_{\rm 2trp,NNN}^{S=0}(k,d)|k,d\rangle\ ,
\end{eqnarray}
\end{subequations}
where 
\begin{subequations}
\begin{eqnarray}
A_{\rm 2trp,NN,weak}^{S=0}(k,d) &=& \sqrt{2}\sum_l a_{l,l+d}^{\rm weak,NN}
\cos\left(k(2l+d)\right) \hspace*{8mm}
\\
A_{\rm 2trp,NNN}^{S=0}(k,d) &=& \\ \nonumber
&& \hspace{-1cm}
\sum_l \sqrt{2}a_{l,l+d}^{\rm L,NNN}\cos\left(k(2l+1/2+d) \right)\ .
\end{eqnarray} 
\end{subequations}
Here $d$ is the distance between the two triplons, 
$a_{l,l+d}^{\rm weak,NN}$ is the amplitude for the creation of two triplons on
 the dimers $l$ and $l+d$ by ${\bf S}_{\rm 0,R}{\bf S}_{\rm 1,L}$, and 
$a_{l,l+d}^{\rm L,NNN}$ is the amplitude for the creation of two triplons on 
the dimers $l$ and $l+d$ by ${\bf S}_{\rm 0,L}{\bf S}_{\rm 1,L}$. It is not necessary
 to calculate $a_{l,l+d}^{\rm R,NNN}$ due to the symmetry 
$a_{l,l+d}^{\rm R,NNN}=a_{-l-d,-l}^{\rm L,NNN}$. The basic unit length $a/2$ is
 again the distance between two neighboring sites. Momentum is 
measured in units of $\pi/(a/2)$.

We first discuss the symmetries of the two observables. 
$R_{\rm NN,weak}^{S=0}$ possesses a reflection symmetry about $k=\pi/2$
for the same reasons as the Hamiltonian. Any mode at $k$ which is 
created by $R_{\rm NN,weak}^{S=0}$ is coupled to the mode at $k+\pi$. In 
addition every mode at $k$ corresponds to the mode at $-k$. Therefore the whole
spectral density will be symmetric about $k=\pi/2$. This symmetry is absent 
for $R_{\rm NNN}^{S=0}$. For $R_{\rm NNN}^{S=0}$ the spectral weight is 
mainly concentrated at small and intermediate momenta  while
it vanishes exactly at $k=\pi$. The latter follows from the fact 
that  at $k=\pi$ the observable creates an odd state with respect to reflection about the axis $1/2$ (see Fig. \ref{fig:Skizze_S0})while a singlet made from two triplets
is always an even state with respect to particle exchange.

In Figs.~\ref{fig:Spectral_D_S0_l03_NN}-\ref{fig:BS_D_S0_l06_NN} the spectral
 densities for $R_{\rm NN,weak}^{S=0}$ are shown and in 
Figs.~\ref{fig:Spectral_D_S0_l03_NNN}-\ref{fig:BS_D_S0_l06_NNN} the 
corresponding densities for $R_{\rm NNN}^{S=0}$. Let us  first
 discuss the results for the nearest neighbor (NN) coupling 
$R_{\rm NN,weak}^{S=0}$ passing then to the results for next-nearest neighbor 
coupling $R_{\rm NNN}^{S=0}$. Finally, the implications 
for Raman spectroscopy and infrared absorption will be assessed.

\subsection{Case $R_{\rm NN,weak}^{S=0}$}

\begin{figure}
  \begin{center}
    \includegraphics[width=\columnwidth]
    {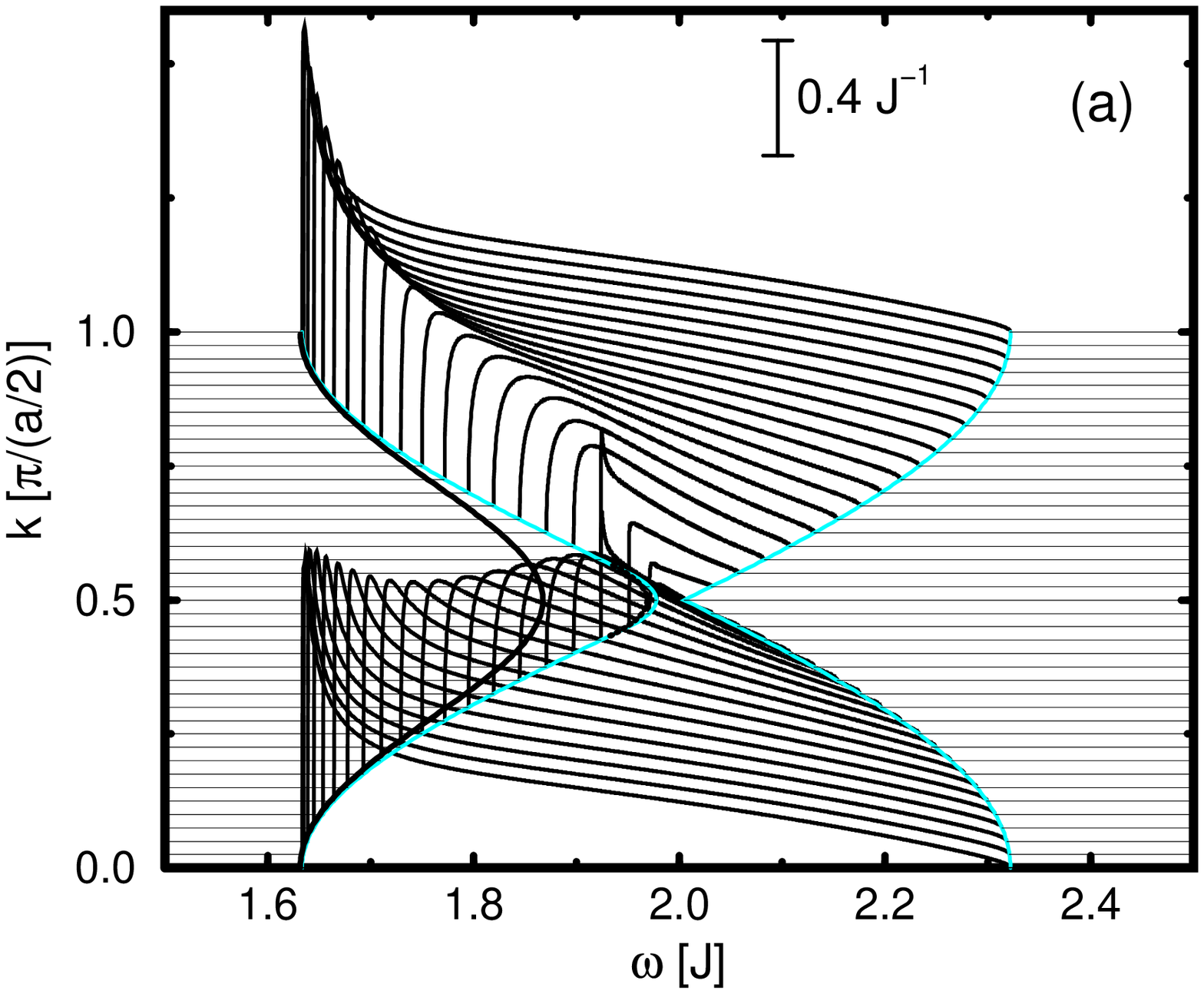}    
    \includegraphics[width=\columnwidth]
    {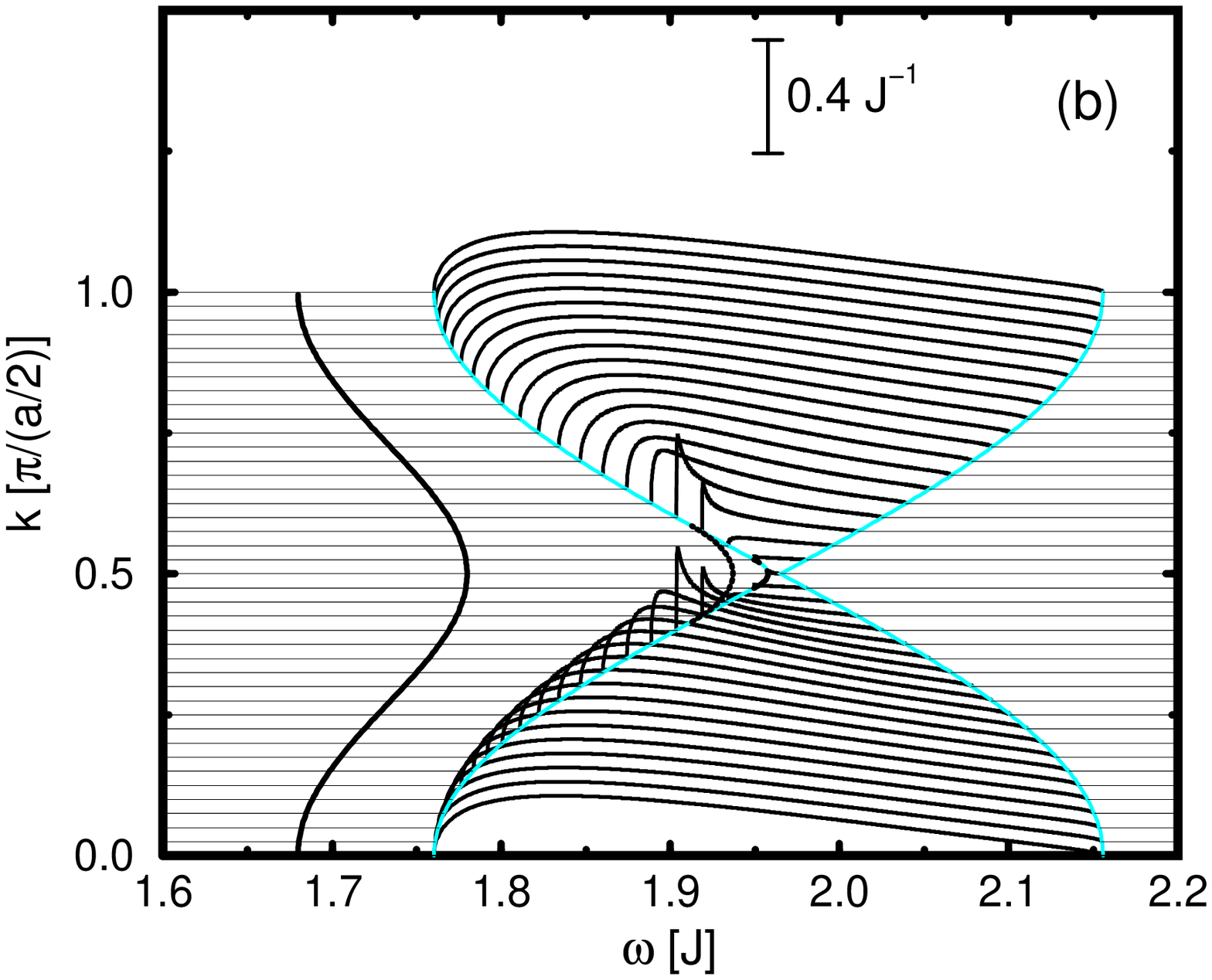} 
    \includegraphics[width=\columnwidth]
    {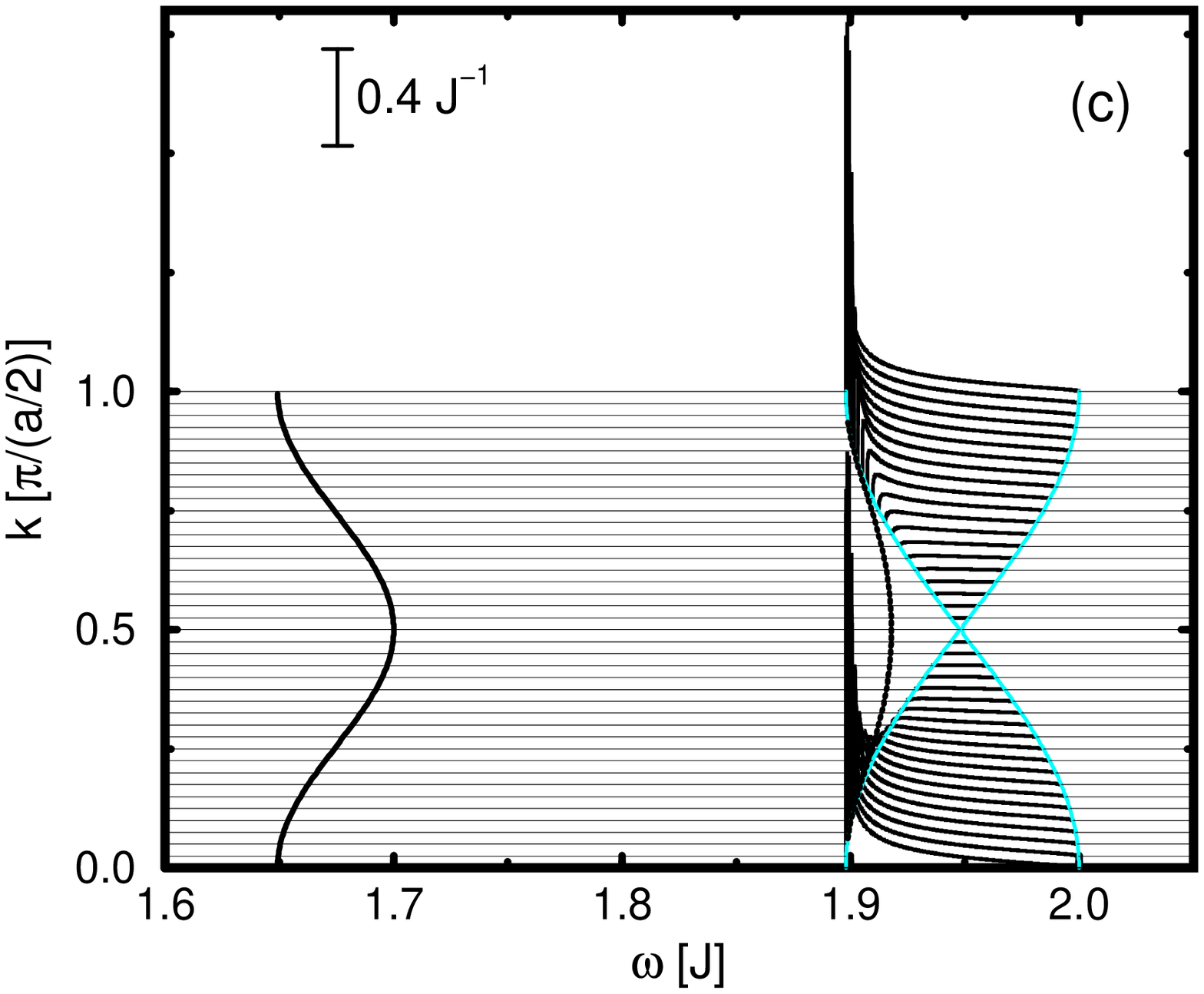} 
    \caption{Two-triplon spectral density $I_2(k,\omega)$ for $R^{S=0}_{\rm NN,weak}$ with
      $\lambda=0.3$ and $\alpha=0.0$ (a), $\alpha=0.25$ (b) and $\alpha=0.5$ 
      (c). 
      Gray lines denote lower and upper bound of the continuum. Black lines 
      indicate dispersion of two-triplon bound states.} 
    \label{fig:Spectral_D_S0_l03_NN}
  \end{center}
\end{figure}

\begin{figure*}
    \includegraphics[width=\textwidth]{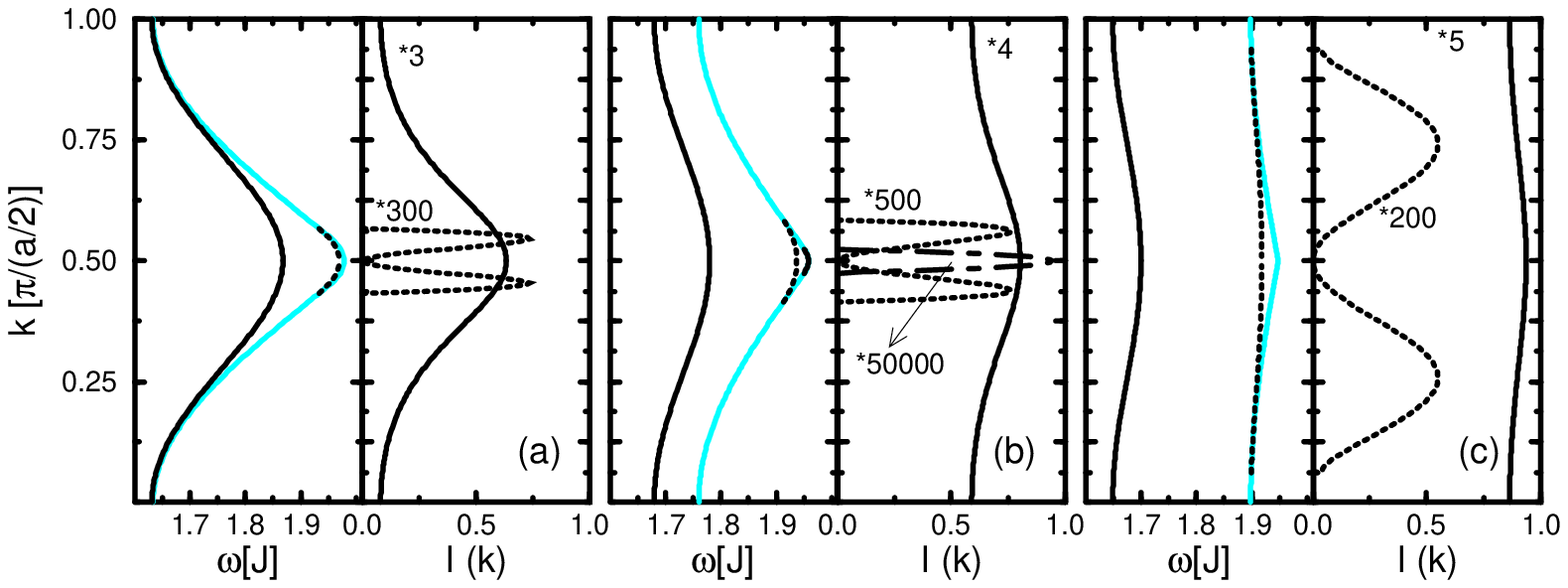} 
    \caption{Two-triplon bound states for $R^{S=0}_{\rm NN,weak}$ with $\lambda=0.3$ and 
      $\alpha=0.0$ (a), $\alpha=0.25$ (b) and $\alpha=0.5$ (c). Left panels
      show the dispersion of the bound states; right panels the 
      spectral weights of the bound states multiplied by the indicated factors. Gray lines denote lower
bound of the continuum.} 
    \label{fig:BS_D_S0_l03_NN}
\end{figure*}

In Fig.~\ref{fig:Spectral_D_S0_l03_NN}a the spectral density for strong 
dimerization ($\lambda=0.3$) and vanishing frustration is depicted. The 
corresponding information about the dispersion and the spectral weight of the
two-triplon bound states is shown in Fig.~\ref{fig:BS_D_S0_l03_NN}a. The same
notation as in the Section for $S=1$ excitations is used. The spectrum is 
symmetric about $k=\pi/2$ due to the inversion symmetry $k\leftrightarrow -k$ 
and $R_{\rm NN,weak}^{S=0}(k+\pi)=R_{\rm NN,weak}^{S=0}(k)$.

We find two $S=0$ two-triplon bound states $S_1$ and $S_2$. The 
triplon-triplon interaction is larger in the total $S=0$ channel than it was 
in the total $S=1$ channel. Therefore the binding energy of the bound states 
is enhanced and the first bound state $S_1$ exists for all momenta in contrast
to the $S=1$ case. In general, the $S=0$ channel is dominated by the
bound states which carry most of the spectral weight. 
This statement applies also to the experimental relevance, see below.

The dispersion of the two-triplon bound state $S_1$ is roughly sinusoidal having three
extrema at momenta $k=\{0;\pi/2;\pi\}$. The binding energy is largest for 
$k=\pi/2$ while it becomes small near momentum zero and $\pi$. The spectral 
weight of $S_1$ is roughly proportional to the binding energy. The second 
singlet two-triplon bound state $S_2$ exists only in a finite interval about
$k=\pi/2$. The spectral weight of $S_2$ vanishes at $k=\pi/2$ and possesses 
two maxima below and above $k=\pi/2$.

\begin{figure}
  \begin{center}
    \includegraphics[width=\columnwidth]
    {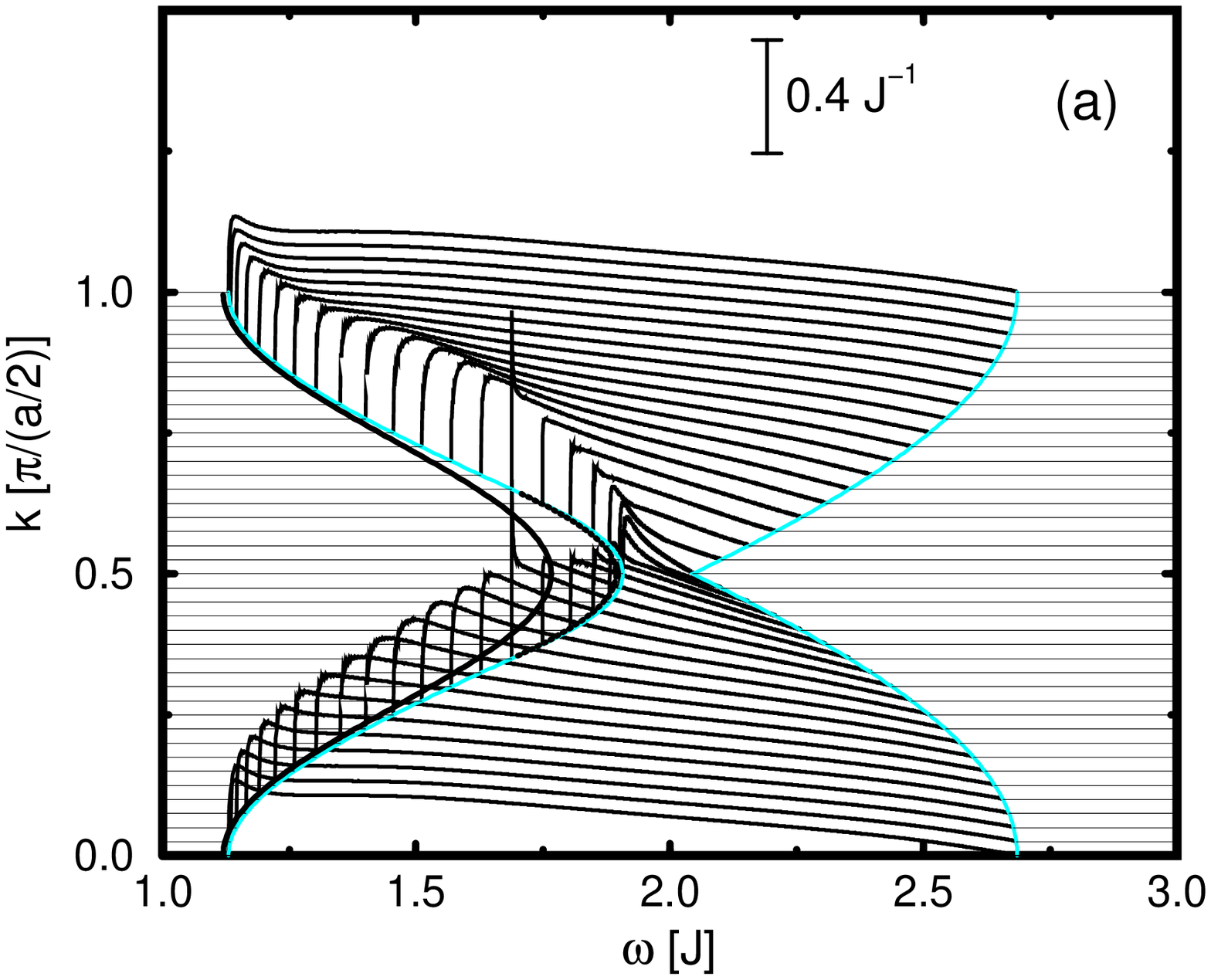}    
    \includegraphics[width=\columnwidth]
    {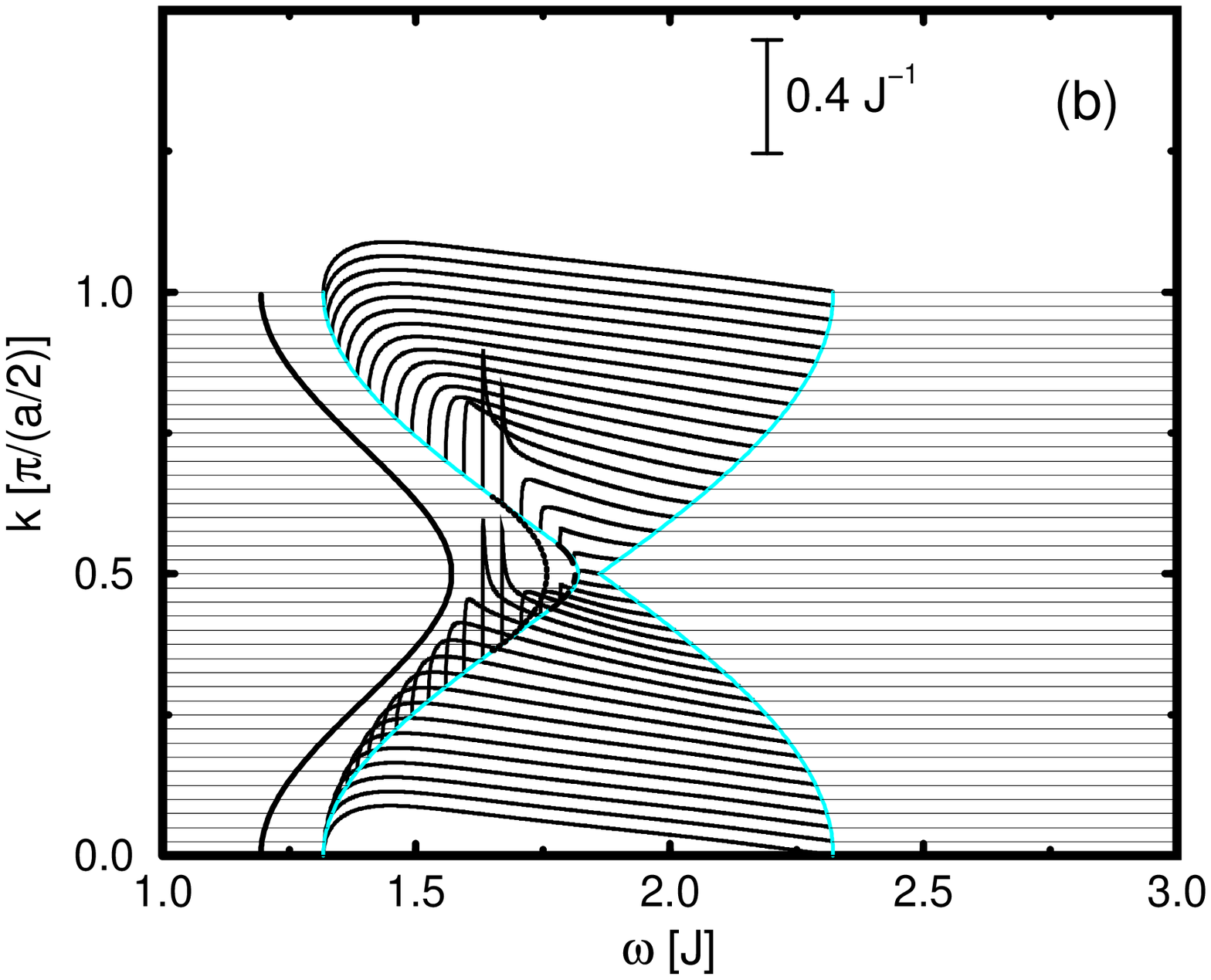}    
    \includegraphics[width=\columnwidth]
    {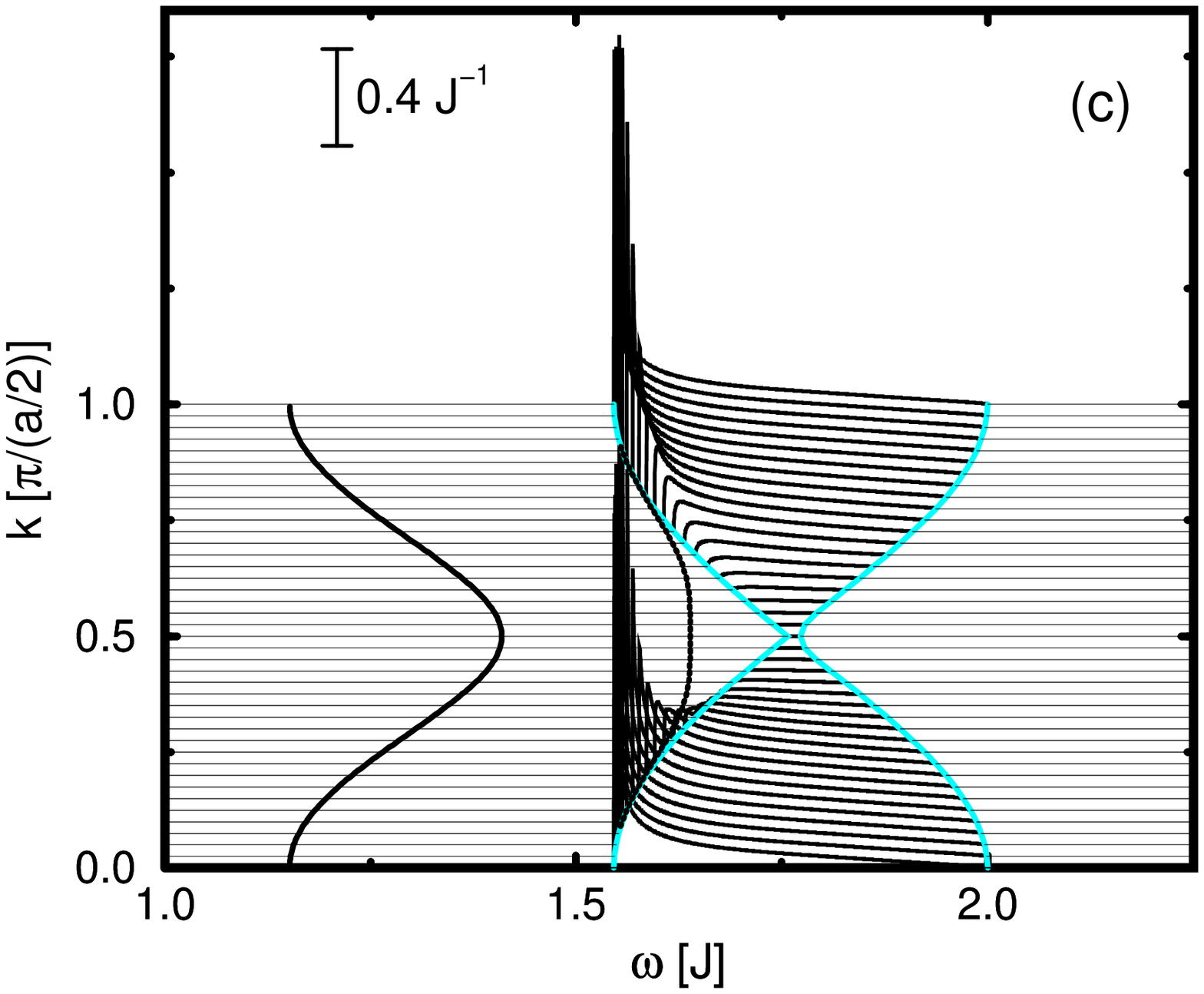} 
    \caption{Two-triplon spectral density $I_2(k,\omega)$ for $R^{S=0}_{\rm NN,weak}$ with
      $\lambda=0.6$ and $\alpha=0.0$ (a), $\alpha=0.25$ (b) and $\alpha=0.5$
      (c).
      Gray lines denote lower and upper bound of the continuum. Black lines 
      indicate dispersion of two-triplon bound states.} 
    \label{fig:Spectral_D_S0_l06_NN}
  \end{center}
\end{figure}

\begin{figure*}
  \includegraphics[width=\textwidth]{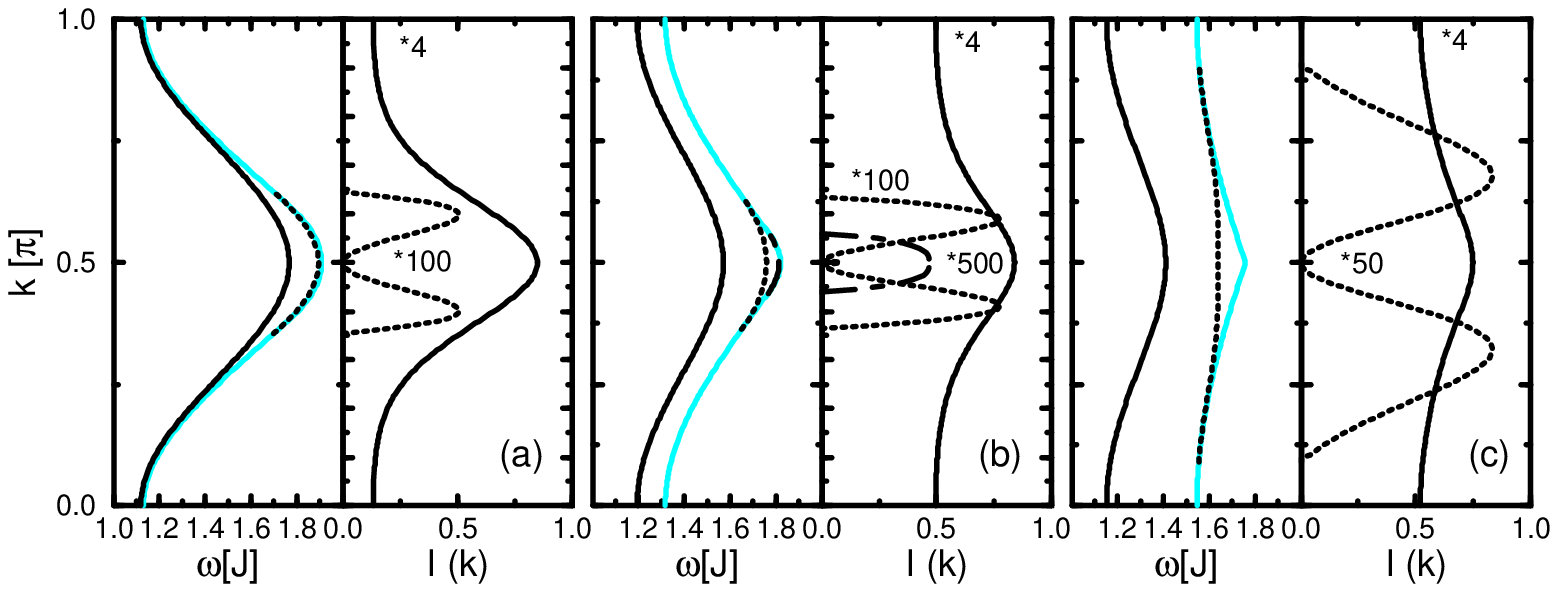} 
  \caption{Two-triplon bound states for $R^{S=0}_{\rm NN,weak}$ with $\lambda=0.6$ and 
    $\alpha=0.0$ (a), $\alpha=0.25$ (b) and $\alpha=0.5$ (c). Left panels show
    the dispersion of the bound states; right panels the spectral 
    weights of the bound states multiplied by the indicated factors.Gray lines denote lower
bound of the continuum. } 
  \label{fig:BS_D_S0_l06_NN}
\end{figure*}

The spectral weight of the two-triplon continuum is concentrated at small 
frequencies. At small and large momenta this effect is enhanced due to the 
vicinity of $S_1$. Lowering the dimerization we see no qualitative changes in
comparison to the case of strong dimerization 
(Fig.~\ref{fig:Spectral_D_S0_l06_NN}a and Fig.~\ref{fig:BS_D_S0_l06_NN}a). 
So we expect that the dispersion of the bound state $S_1$ is degenerate with
the lower band edge of the two-triplon continuum inducing
a square root divergence. This expectation is supported also by
numerical results for the second breather\cite{bouze98a}.

For clarity, we like to emphasize again that one must clearly distinguish
the the case of zero dimerization and the case of small, but finite,
dimerization. For zero dimerization bosonization predicts a $1/\omega$ 
divergence at $k=\pi$ which becomes $(\omega-\omega_0)^{-1/2}$ close to 
$k=\pi$\cite{halda82b,cross79}. This has been used for instance
in the empirical calculation of Lorenzana and Eder \cite{loren97c}.
For small, but finite, dimerization the sine-Gordon model prediction
of a square root behavior without divergence applies to the critical
frustration and in the region around this value. For other values of
the frustration a breather may coincide with the lower band edge
implying a square root divergence.

In Figs.~\ref{fig:Spectral_D_S0_l03_NN}b-c the spectral density at finite
frustration for strong dimerization is shown. At $\alpha=0.25$ we find three 
bound states $S_1$, $S_2$ and $S_3$. The binding energy of $S_1$ increases 
drastically on turning on the frustration, especially at small and large 
momenta. The spectral weight increases in a similar fashion for these momenta.
The third two-triplon bound state $S_3$ exists merely in a very small region about 
$k=\pi/2$. The binding energy and the spectral weight are tiny. The 
spectral weight has a maximum at $k=\pi/2$.

The two-triplon continuum for $\alpha=0.25$ does not show much structure. This 
is a consequence of the fact that at almost all momenta no bound 
state is close to the lower band edge of the continuum. Decreasing the 
dimerization no qualitative changes are seen 
(see Fig.~\ref{fig:Spectral_D_S0_l06_NN}b).

At $\alpha=0.5$ we detect two bound states. The dispersion of $S_1$ becomes 
flatter which again holds also for the spectral weight distribution. The 
biggest change can be seen in $S_2$. This bound state exists for almost all
momenta in contrast to the cases $\alpha=\{0;0.25\}$. In the regions close
to $k=0$ and symmetrically close to $k=\pi$, the bound state $S_2$ does not 
exist, but it can be thought to lie just above the lower band edge implying 
an almost divergent behavior of the two-triplon continuum.

Smirnov\cite{smirn92,gogol98} showed that the corresponding spectral density 
of the sine-Gordon model displays a square root behavior at the
lower band edge. This applies to the $S=0$ channel of 
the frustrated spin chain at $\alpha\leqq\alpha_{0,c}$ if
the marginal term $D\cos(4\Phi)$ of Eq.~(\ref{H_FT}) is neglected.
This neglect is quantitatively justified for $\alpha=\alpha_{0,c}$. 
Indeed, our results clearly show a square root behavior for 
$\alpha=0.25$. As for the $S=1$ case, we find that the 
predictions of the sine-Gordon model for the physics of the spin
chain are verified for $\alpha=\alpha_{0,c}$. 
For other values, notably $\alpha=0$ and $\alpha=0.5$, we find 
square root divergences or strong tendencies towards square root divergences.
Again, such divergent behavior results from the vicinity of bound states,
here in the $S=0$ sector.

\subsection{Case $R_{\rm NNN}^{S=0}$}

\begin{figure}
  \begin{center}
    \includegraphics[width=\columnwidth]
    {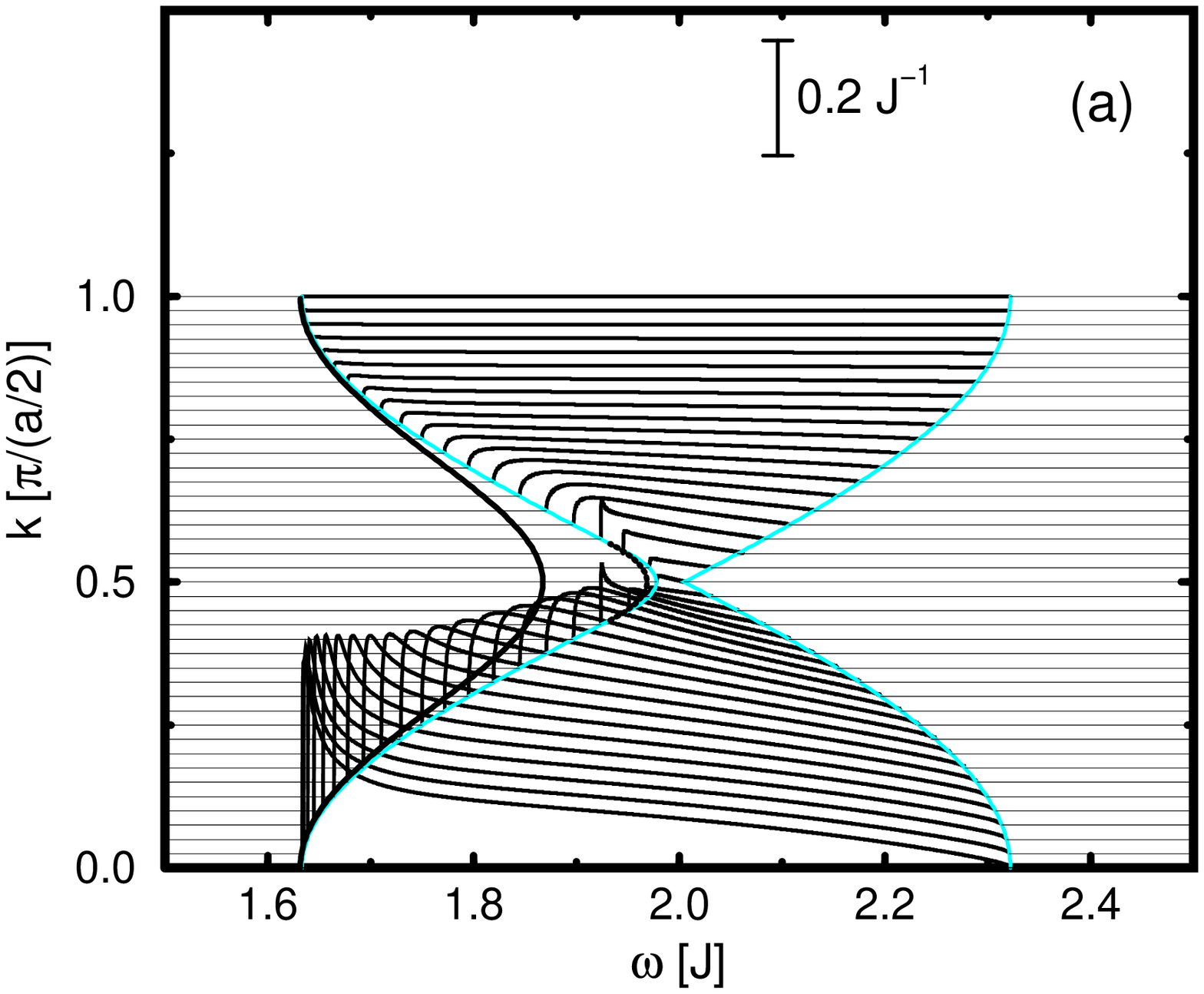} 
    \includegraphics[width=\columnwidth]
    {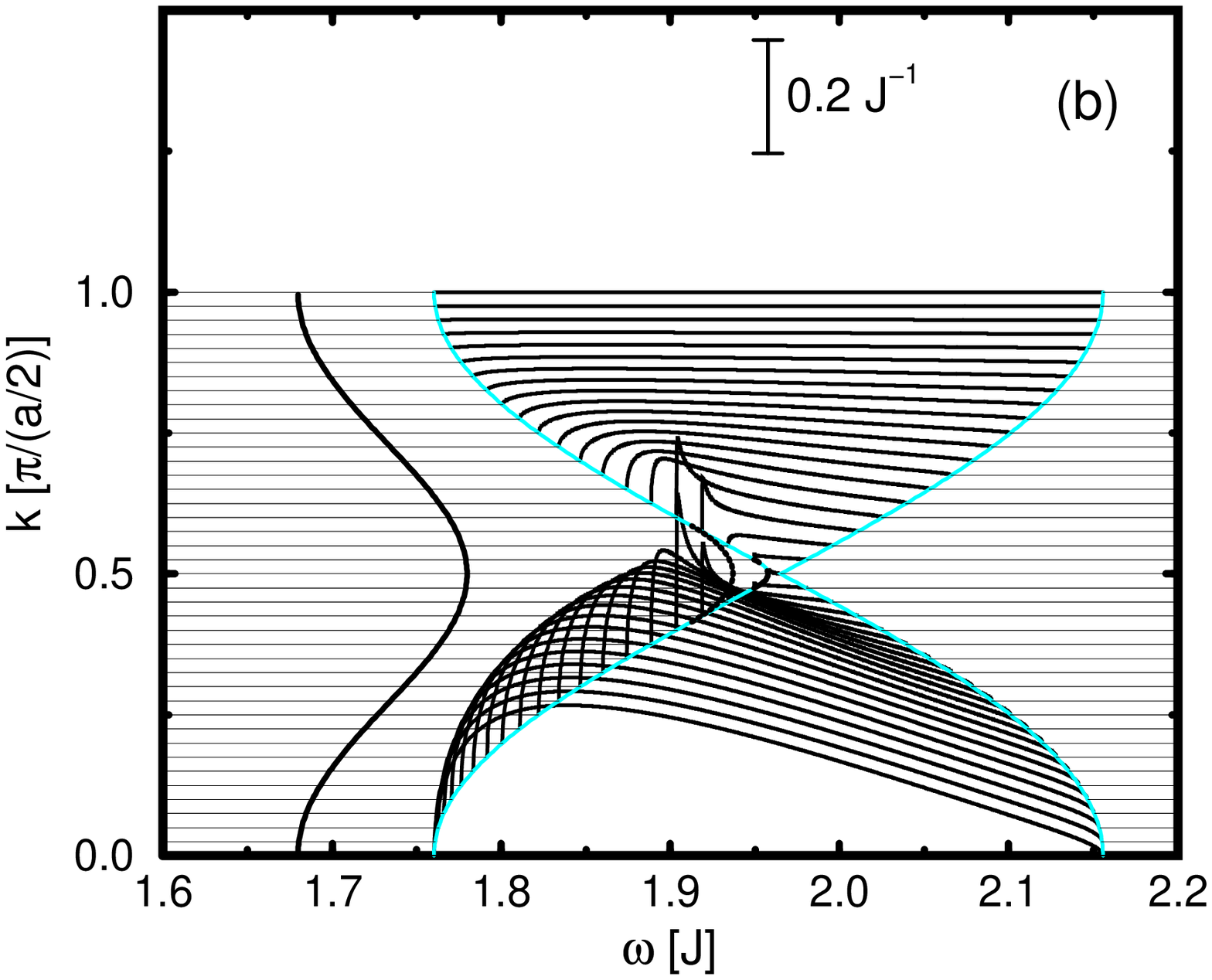} 
    \includegraphics[width=\columnwidth]
    {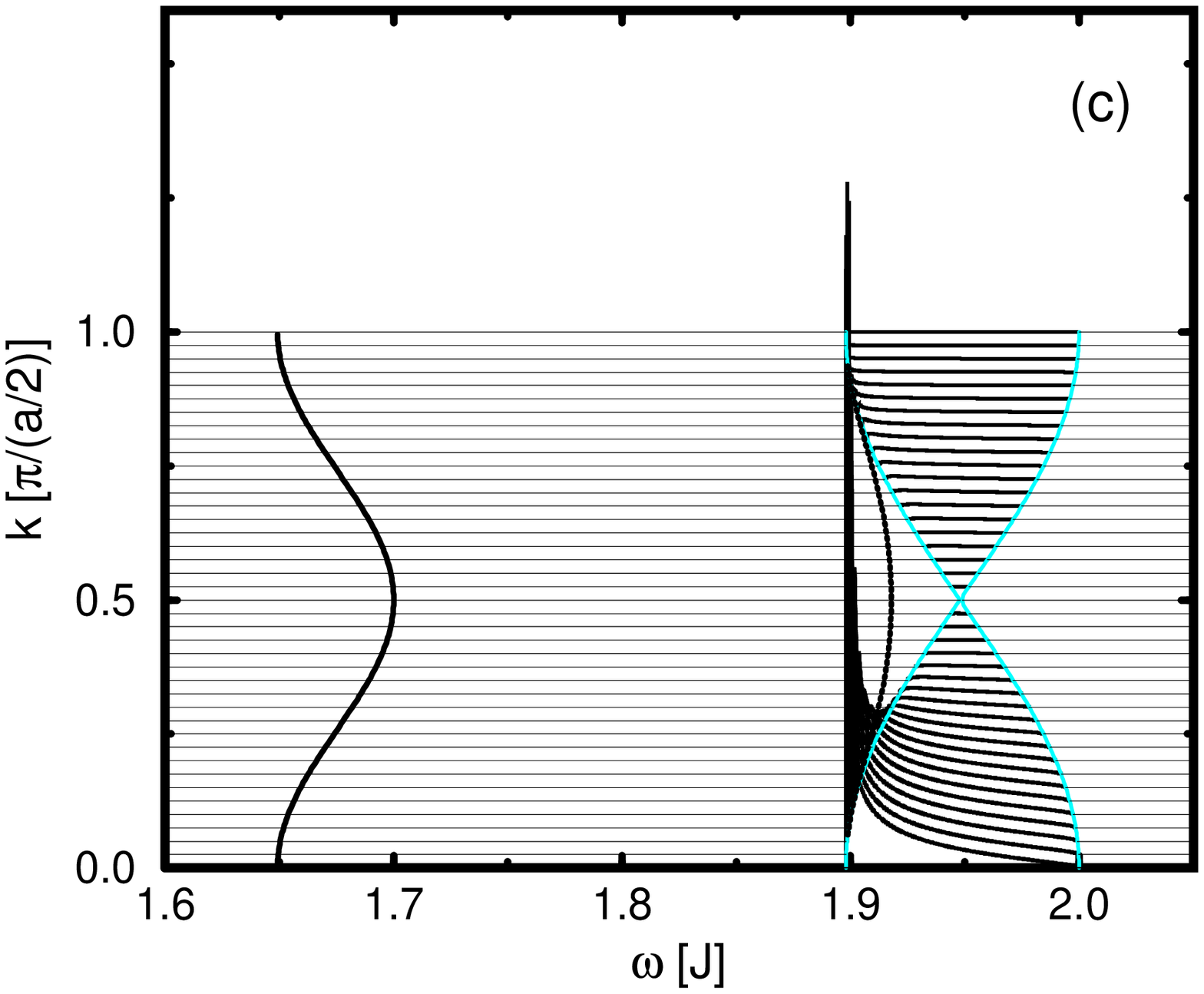} 
    \caption{Two-triplon spectral density $I_2(k,\omega)$ for $R^{S=0}_{\rm NNN}$ with 
      $\lambda=0.3$ and $\alpha=0.0$ (a), $\alpha=0.25$ (b) and $\alpha=0.5$ 
      (c).
      Gray lines denote lower and upper bound of the continuum. Black lines 
      indicate dispersion of two-triplon bound states.} 
    \label{fig:Spectral_D_S0_l03_NNN}
  \end{center}
\end{figure}

\begin{figure*}
  \includegraphics[width=\textwidth]{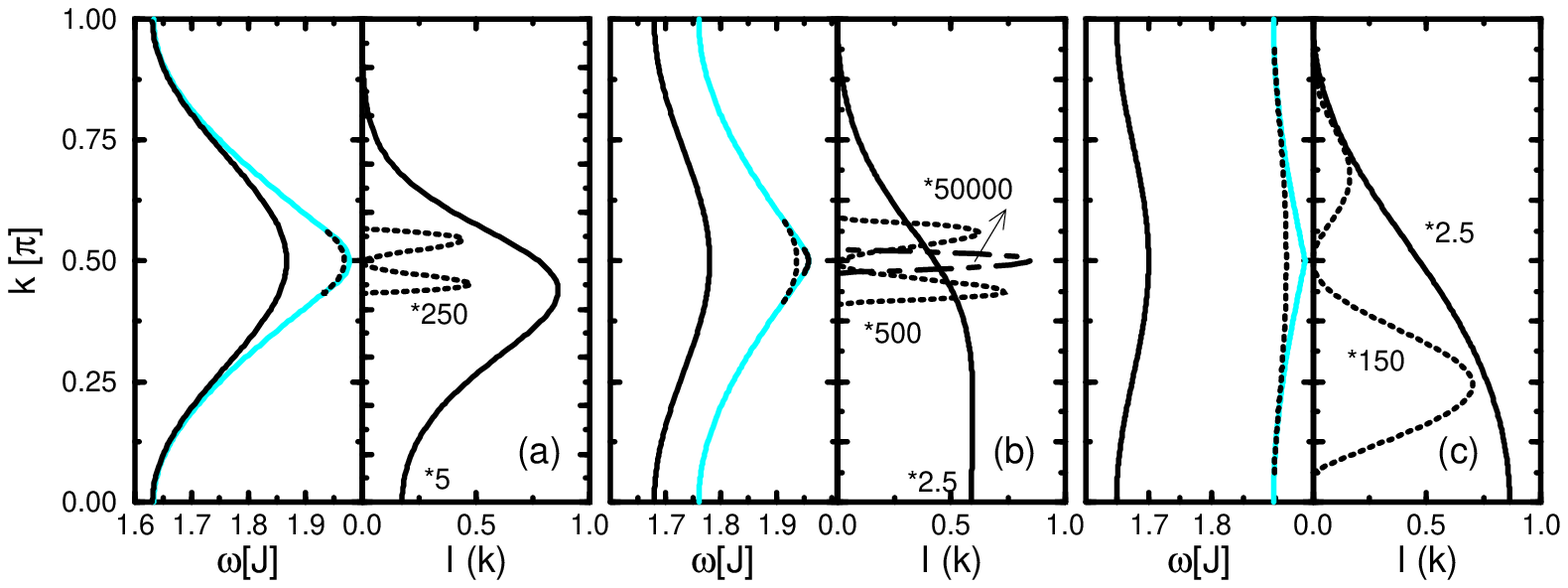}
  \caption{Two-triplon bound states for $R^{S=0}_{\rm NNN}$ with $\lambda=0.3$ and 
    $\alpha=0.0$ (a), $\alpha=0.25$ (b) and $\alpha=0.5$ (c). Left panels 
    show the dispersion of the bound states; right panels the 
    spectral weights of the bound states multiplied by the indicated 
    factors. Gray lines denote lower
bound of the continuum.} 
  \label{fig:BS_D_S0_l03_NNN}
\end{figure*}

\begin{figure}
  \begin{center}
    \includegraphics[width=\columnwidth]
    {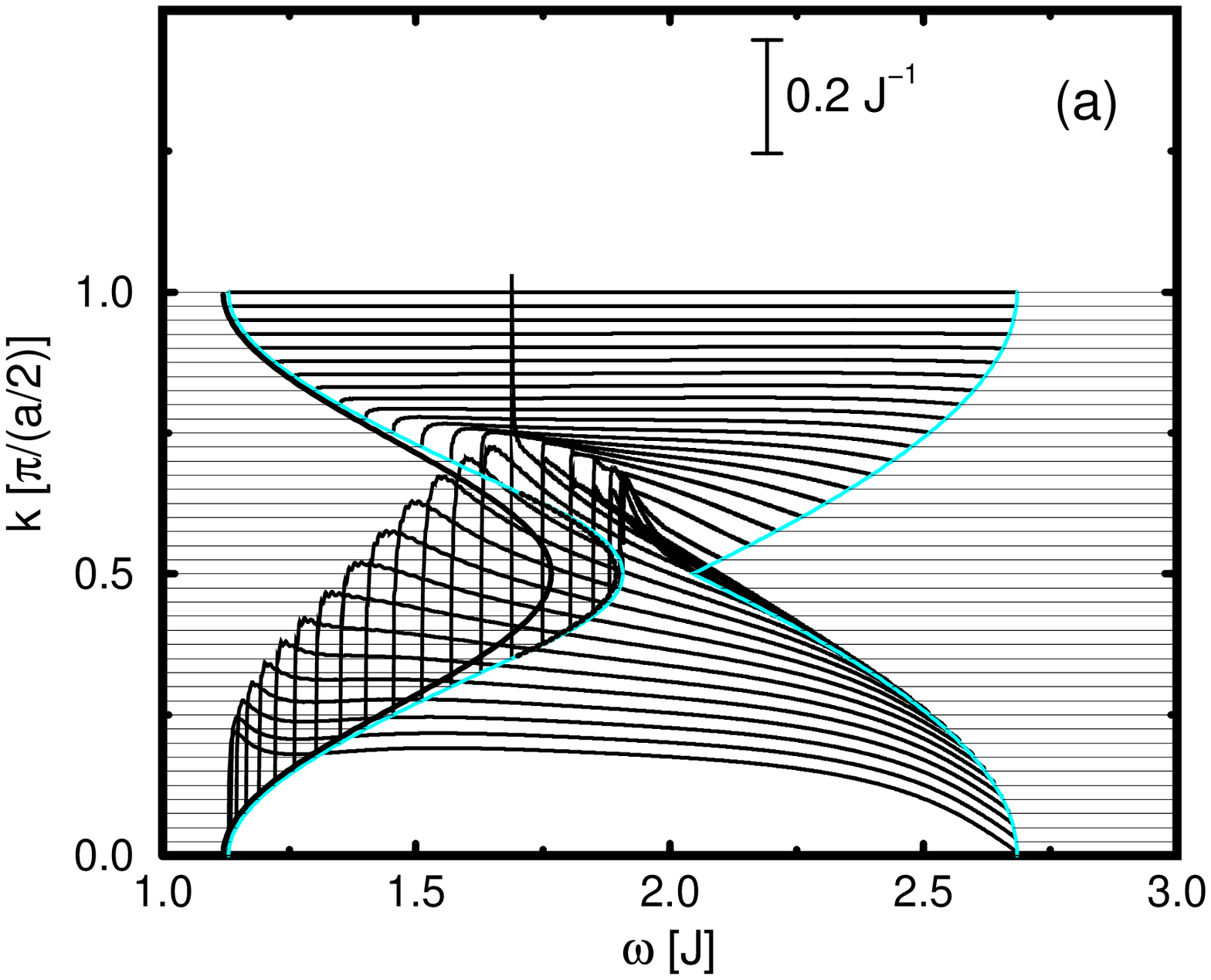} 
    \includegraphics[width=\columnwidth]
    {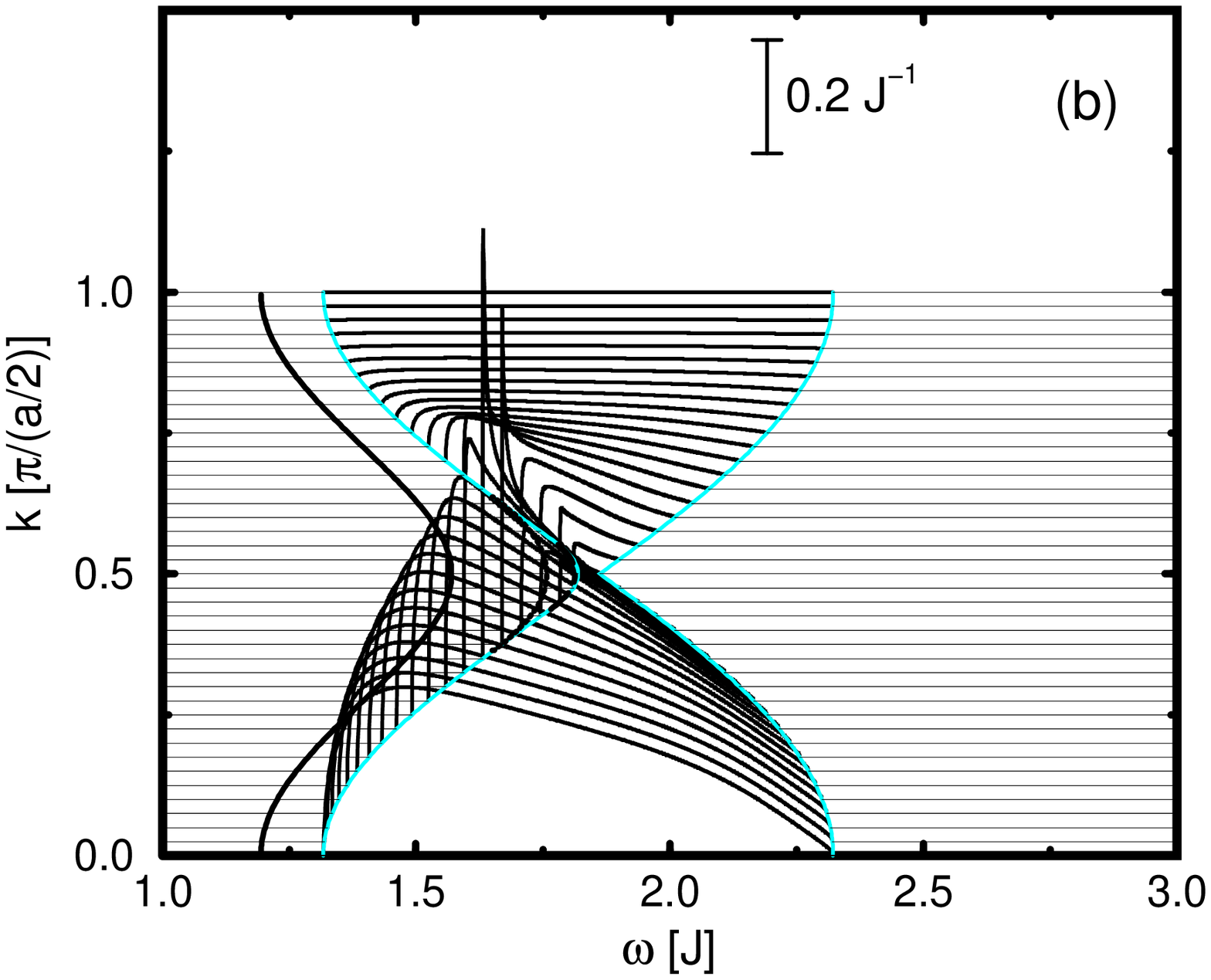} 
    \includegraphics[width=\columnwidth]
    {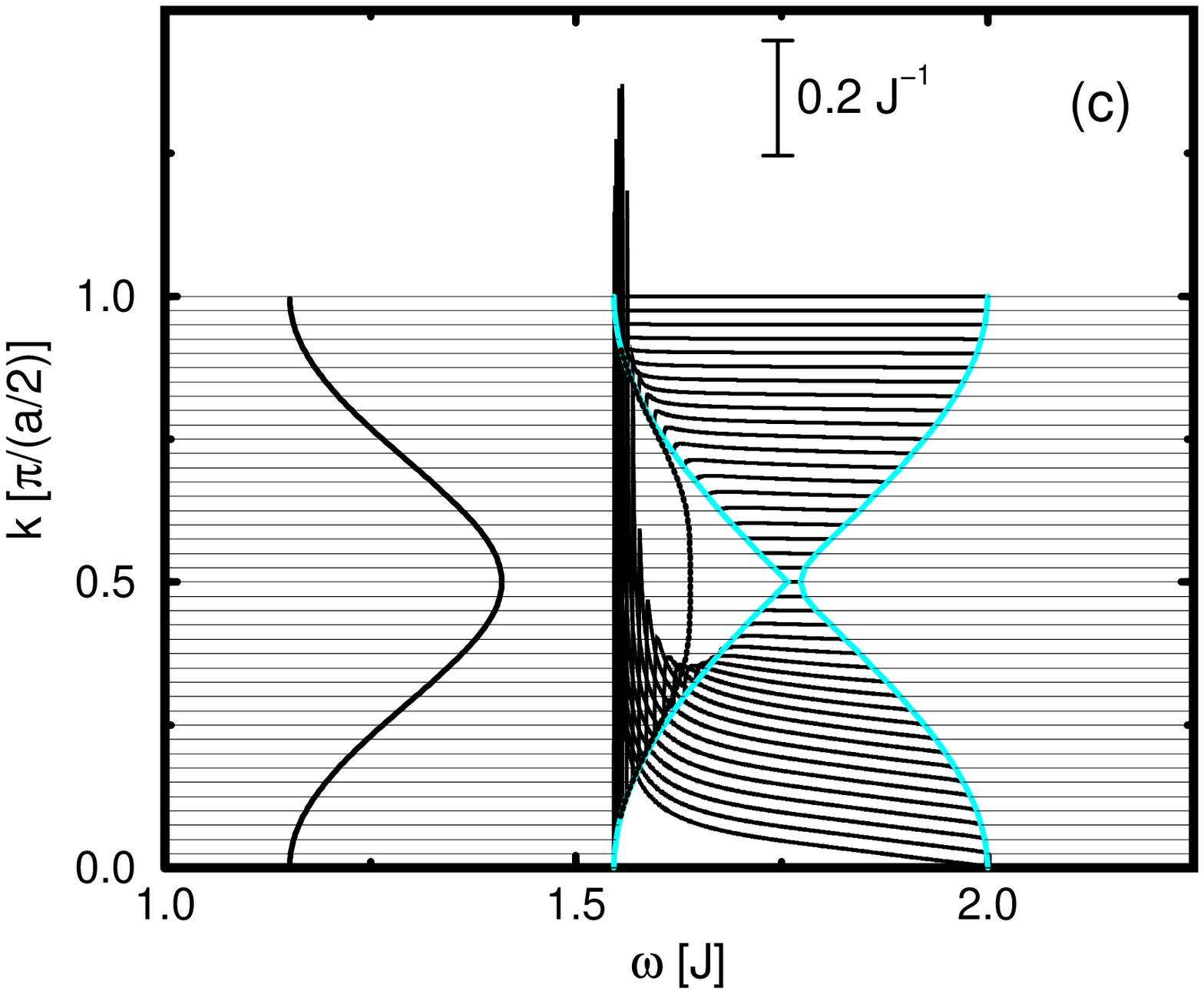} 
    \caption{Two-triplon spectral density $I_2(k,\omega)$ for $R^{S=0}_{\rm NNN}$ with 
      $\lambda=0.6$ and $\alpha=0.0$ (a), $\alpha=0.25$ (b) and $\alpha=0.5$ (c). 
      Gray lines denote lower and upper bound of the continuum. Black lines 
      indicate dispersion of two-triplon bound states.} 
    \label{fig:Spectral_D_S0_l06_NNN}
  \end{center}
\end{figure}

\begin{figure*}
  \includegraphics[width=\textwidth]{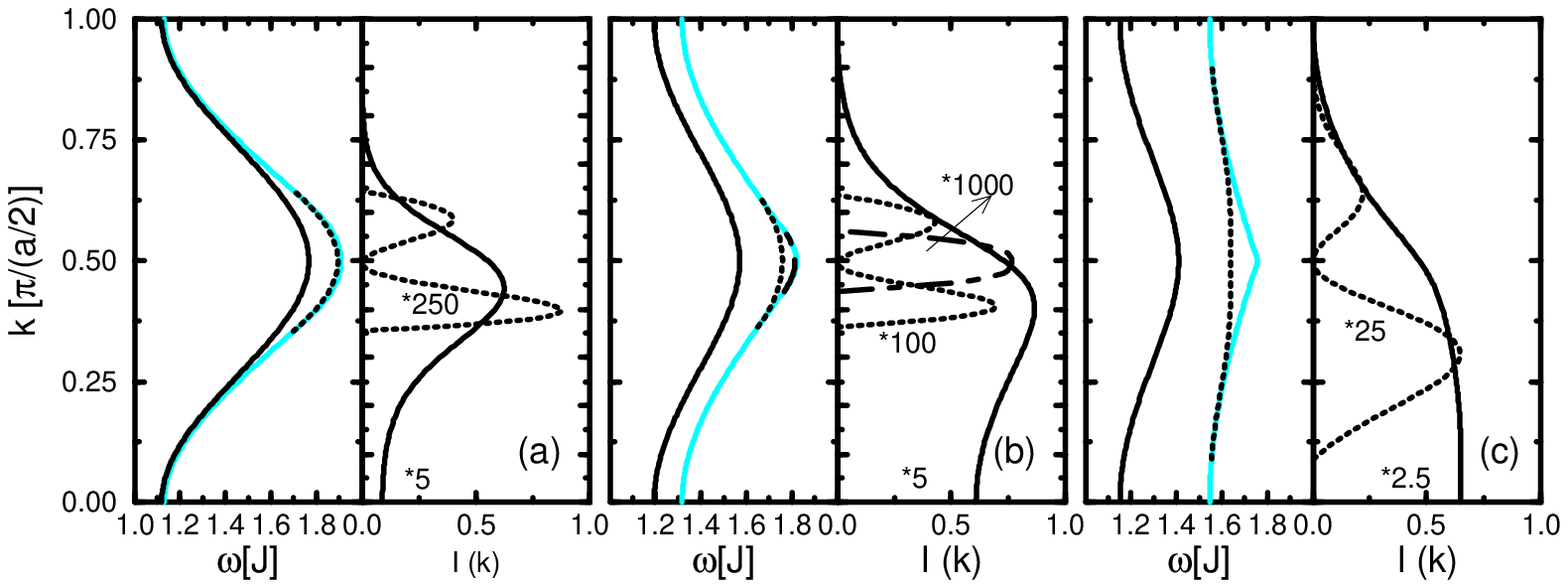}
  \caption{Two-triplon bound states for $R^{S=0}_{\rm NNN}$ with $\lambda=0.6$ and 
    $\alpha=0.0$ (a), $\alpha=0.25$ (b) and $\alpha=0.5$ (c). Left panels show
    the dispersion of the bound states; right panels the spectral 
    weights of the bound states multiplied by the indicated factors. Gray lines denote lower
bound of the continuum.}
  \label{fig:BS_D_S0_l06_NNN}
\end{figure*}

In Figs.~\ref{fig:Spectral_D_S0_l03_NNN} and \ref{fig:Spectral_D_S0_l06_NNN} 
the spectral densities of the observable $R_{\rm NNN}^{S=0}$ for various values 
of dimerization and frustration are shown. The information on the singlet 
two-triplon bound states is plotted in Figs.~\ref{fig:BS_D_S0_l03_NNN} and 
\ref{fig:BS_D_S0_l06_NNN}. All considerations concerning the energetic 
properties of the $S=0$ channel are the same as for $R_{\rm NNN}^{S=0}$ and 
need not be discussed again in this section.

We concentrate on the spectral differences between the two observables. The 
spectral weight is suppressed for large momenta due to symmetry reasons. This
can be clearly seen for the two-triplon continuum and the spectral weight of 
the two-triplon bound states. Therefore only momenta $k\in [0,\pi/2]$ carry 
significant spectral weight. In this region we do not find qualitative changes
to the results for of $R_{\rm NN,weak}^{S=0}$.

The most important difference is a change of the spectral weight distribution of 
$S_1$. At $\alpha=0$ the spectral weight has a maximum at $k=\pi/2$ which is 
similar to the case of $R_{\rm NN,weak}^{S=0}$. Finite frustration 
shifts the maximum to $k=0$. Close to the critical frustration
the spectral weight is almost constant for $k\in[0,\pi/2]$. At $\alpha=0.5$ 
the spectral weight is a monotonically decreasing function (from $k=0$ to 
$k=\pi$).

\subsection{Raman Spectroscopy}

The dominant observable for magnetic light scattering (Raman response) using 
the standard Fleury-Loudon scattering theory\cite{fleur68,shast90b} is
\begin{equation}
 R_{\rm Raman} = \sum_i \left( R^{S=0}_{\rm loc,NN}(i) +
\beta R^{S=0}_{\rm loc,NNN}(i)\right) ,
\end{equation} 
where the sum runs over all spins. The Raman response is therefore the $k=0$ 
contribution to the spectral density we have discussed in the last section.
We focus here on the case of next-nearest neighbor coupling which is the 
leading process in the case of a uniform Heisenberg chain without 
frustration. 

\begin{figure*}
    \includegraphics[width=\textwidth]{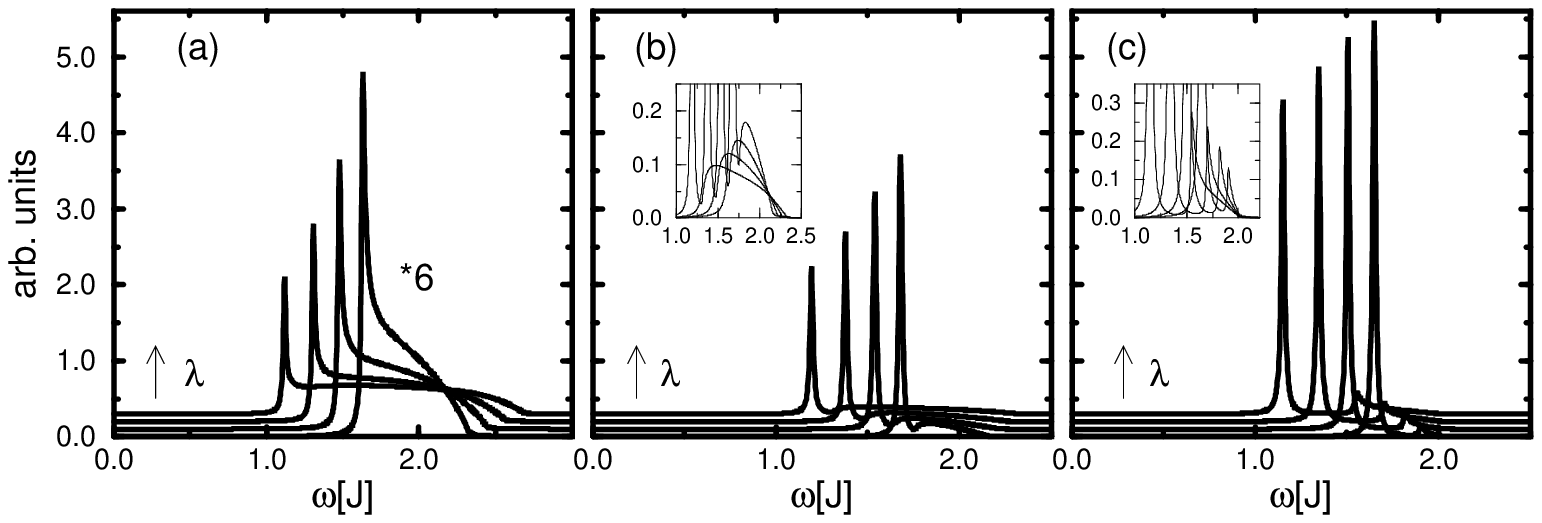} 
    \caption{Raman line shape for $R^{S=0}_{\rm NNN}$ with additional 
broadening $\Gamma=0.01$: (a) $\alpha=0$ (b) $\alpha=0.25$ (c) $\alpha=0.5$. 
In each picture, curves for $\lambda=\{ 0.3;0.4;0.5;0.6\}$ are shown, shifted
for increasing $\lambda$ in $y$-direction. The Raman 
response for $\alpha=0$ is multiplied by 6. The insets in (b) and (c)
zoom on the continua.} 
    \label{fig:R_S0_a0_NNN} 
\end{figure*}

In Fig.~\ref{fig:R_S0_a0_NNN} the Raman response for next-nearest coupling is
shown at zero frustration (a), close to critical frustration $\alpha=0.25$ (b)
and for $\alpha=0.5$ (c). In each graph the spectrum is shown for 
$\lambda=\{0.3;0.4;0.5;0.6\}$. In these figures, a broadening of 
$\Gamma=0.01$ is used and the spectra are shifted in $y$-direction for clarity.
The spectral densities for $\alpha=0$ are multiplied by 6.

For the dimerizations considered here, the spectra are dominated by the first
$S=0$ two-triplon bound state $S_1$. This dominance is enhanced by the 
frustration. In Fig.~\ref{fig:R_S0_a0_NNN}a the case of vanishing 
frustration is shown. Due to the finite broadening and the small binding 
energy of the bound state there is no separation of the two-triplon bound 
state and the two-triplon continuum. An increase of $\lambda$ reduces the 
weight of the bound state and gives rise to a broad featureless 
continuum.

In Fig.~\ref{fig:R_S0_a0_NNN}b and \ref{fig:R_S0_a0_NNN}c the results for 
finite frustration are plotted. The binding energy of the bound state $S_1$
is enhanced and one can clearly separate the contribution of the bound state
$S_1$ and the continuum. The spectral weight of the two-triplon continuum is 
very small.

\subsection{IR-Absorption}

In this section we apply our results to phonon-assisted infrared absorption of
 magnetic excitations\cite{loren95a,loren95b}. This technique allows to study 
the spin-spin correlation function by measuring the optical conductivity. The 
direct absorption of two magnetic excitations is generically not allowed due
 to inversion symmetry. However, this selection rule can be broken by 
 simultaneously exciting a phonon. The leading infrared-active magnetic 
absorption is a two-triplon-plus-phonon process\cite{loren95a,loren95b}. 
Due to the momentum of the excited phonon, the magnetic spectra $I(k,\omega)$ 
have to be integrated over all momenta weighted with a phonon-specific form 
factor.

The absorption spectra are sensitive to the $S=0$ two-triplon bound states.
Especially the extrema yield prominent van-Hove singularities in the 
density of states which can be identified in experiment. In this way,
the first experimental evidence for the two-triplon bound state in 
cuprate spin ladder systems\cite{windt01} was possible. We therefore expect 
interesting line shapes in the optical absorption also for  
dimerized and frustrated spin chain systems.

The absorption coefficient is given by\cite{loren97c}
\begin{equation}
 \label{Eq:IR}
 a(\omega)=a_0\omega I^{\rm IR}(\omega-\omega_0).
\end{equation}
Here $a_0$ is a constant depending on the material and $\omega_0$ is the
 phonon frequency. The phonon is considered to be local and without 
dispersion. The function $I^{\rm IR}$ is given by
\begin{equation}
 I^{\rm IR}(\omega) = 16\pi \sum_k \sin^4 (k/2) I(k,\omega).
\end{equation}     
The specific form factor given is strictly valid only for a uniform Heisenberg 
chain. It was successfully used to explain the optical absorption in uniform
cuprate spin chains\cite{loren97c,suzuu96}. We use the same form factor also for the 
dimerized and frustrated chain in order to explore the general features of the
optical conductivity and to compare the line shapes at finite dimerization 
with the line shapes at zero dimerization. In a detailed analysis of
experimental data one must analyze which phonons 
are involved and which specific form factors matter.

\begin{figure*}
    \includegraphics[width=\textwidth]{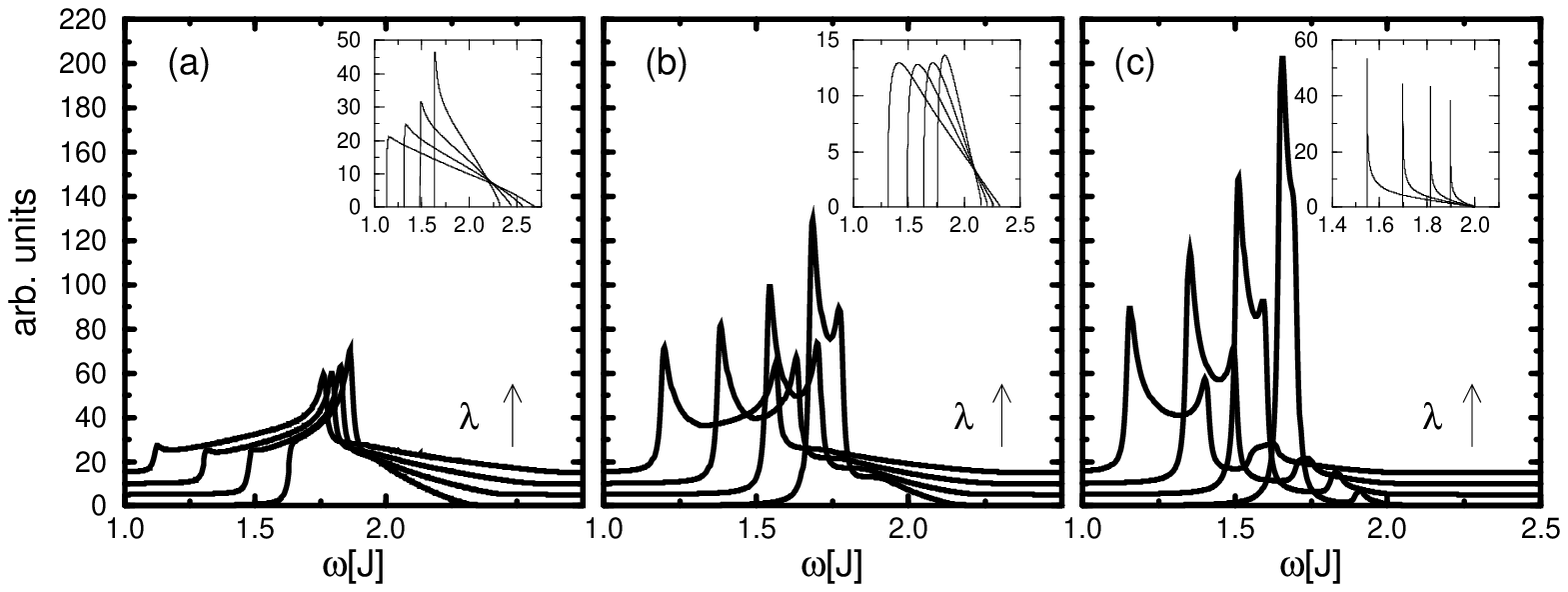} 
    \caption{Optical absorption for $R^{S=0}_{\rm NN,weak}$ with additional 
broadening $\Gamma=0.01$: (a) $\alpha=0.0$ (b) $\alpha=0.25$ (c) $\alpha=0.5$.
 In each picture, curves for $\lambda=\{ 0.3;0.4;0.5;0.6\}$ are shown. Insets:
 Contribution of the two-triplon continuum without broadening.} 
    \label{fig:IR_S0_NN}
\end{figure*}
\begin{figure*}
    \includegraphics[width=\textwidth]{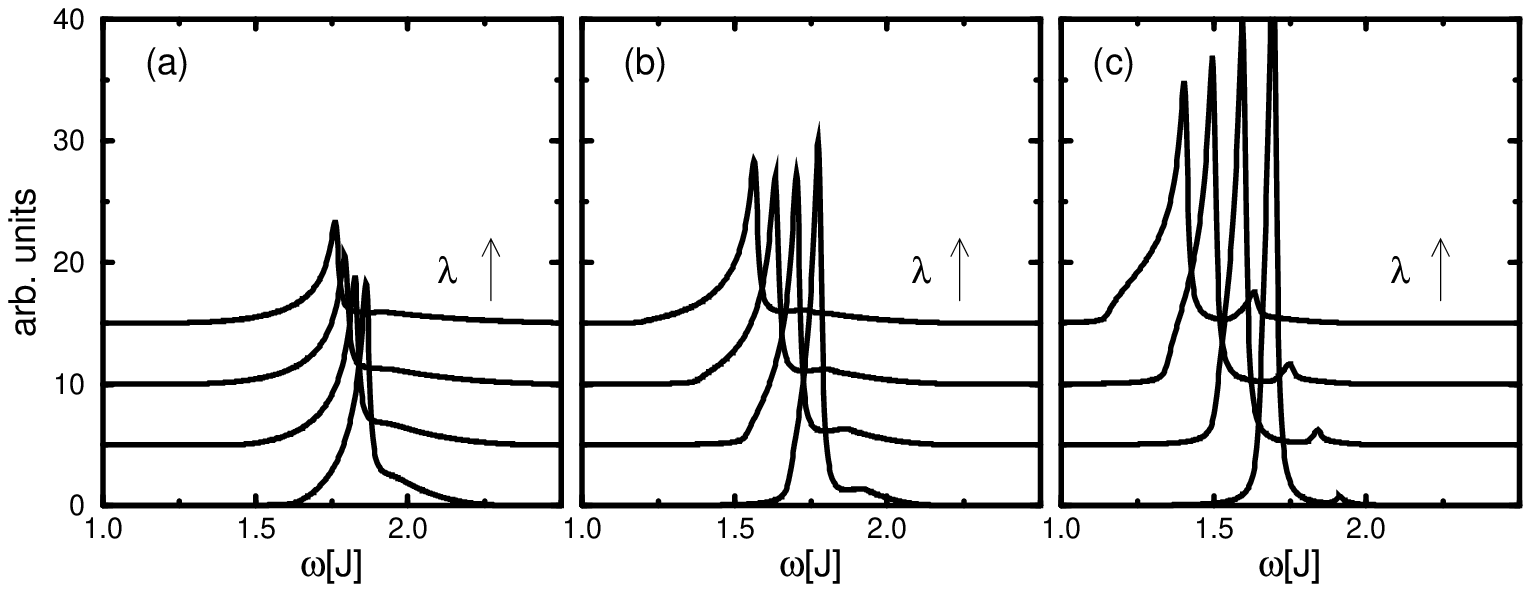} 
    \caption{Optical absorption for $R^{S=0}_{\rm NNN}$ with additional 
broadening $\Gamma=0.01$: (a) $\alpha=0.0$, (b) $\alpha=0.25$, (c) $\alpha=0.5$.
 In each picture, curves for $\lambda=\{ 0.3;0.4;0.5;0.6\}$ are shown.} 
    \label{fig:IR_S0_NNN}
\end{figure*}

In Fig.~\ref{fig:IR_S0_NN}a-c and Fig.~\ref{fig:IR_S0_NNN}a-c the optical 
absorption $a (\omega)$ for various dimerization and frustration is shown
 for $R^{S=0}_{\rm NNN}$ and $R^{S=0}_{\rm NN,weak}$. Here $a_0$ is set 
to one and $\omega_0$ is set to zero. The spectra are plotted with a broadening 
of $\Gamma=0.01$ which is a reasonable value in view of experimental resolutions. 
In the insets we show the contribution of the two-triplon continuum without the 
broadening to highlight the shape of the continuum contributions and to 
distinguish it from contributions of the bound states.

The phonon form factor favors large momenta while it reduces the contribution
 of small momenta. Hence the discussion of the spectral densities implies that 
 $R^{S=0}_{\rm NN,weak}$ is more relevant
 than $R^{S=0}_{\rm NNN}$ for $a(\omega)$. This can also be seen in the 
absolute heights of the spectra in Fig.~\ref{fig:IR_S0_NN} and 
Fig.~\ref{fig:IR_S0_NNN}. In addition, we expect that the nearest neighbor 
coupling is stronger than the next-nearest neighbor one because exchange 
processes of longer range will generically be less important.

We start our discussion with $R^{S=0}_{\rm NN,weak}$. As stated above the first 
$S=0$ two-triplon bound state $S_1$ carries most of the spectral weight 
for all momenta. Hence, it is of crucial importance for the optical absorption. In 
Figs.~\ref{fig:BS_D_S0_l03_NN} and \ref{fig:BS_D_S0_l06_NN}, we show that the
 dispersion $\omega_{\rm bound}(k)$ of the bound state $S_1$ possesses three 
extrema at $k=\{0,\pi/2,\pi\}$. So we obtain three van-Hove singularities in 
$I^{\rm IR}(\omega)$. The spectral density is symmetric about $k=\pi/2$ so that
two van-Hove singularities coincide and there are
two peaks resulting from the bound state $S_1$ in the optical conductivity. 
The weight of the minimum at $k=0$ is suppressed by the phonon form factor.
This implies that the regions about $k=\pi/2$ and $k=\pi$ dominate. 
 
In Fig.~\ref{fig:IR_S0_NN}a the optical absorption for a dimerized chain 
($\lambda=\{ 0.3,0.3,0.5,0.6\}$) without frustration is depicted. The spectra are
shifted in $y$-direction for clarity. The line shape is dominated by a small 
peak at low energies, a sharp peak at intermediate energies and a broad 
structure at high energies. The first two features are mainly produced by the 
above mentioned van-Hove singularities resulting from the extrema of the 
bound state dispersion of $S_1$. The second peak is dominant because the 
spectral weight has a maximum for $k=\pi/2$, see Fig.~\ref{fig:BS_D_S0_l03_NN}a and \ref{fig:BS_D_S0_l06_NN}a. For increasing $\lambda$ this peak looses 
intensity while the first peak becomes more pronounced. The latter effect is 
due to the increasing binding energy of $S_1$ at $k=\pi$.

For strong dimerization the feature at low energies, which is more like a 
shoulder than like peak, is an effect of the two-triplon continuum (inset 
Fig.~\ref{fig:IR_S0_NN}a). The second bound state $S_2$ is of no greater 
relevance for the optical absorption because it has zero spectral weight for 
$k=\pi/2$ which is the only extremum of the bound state dispersion. In addition,
the binding energy is very small without frustration and so is the 
corresponding spectral weight.

Lorenzana and Eder\cite{loren97c} calculated the two-spinon-plus-phonon 
contribution to the optical absorption for a uniform Heisenberg chain. The 
line shape consists mainly of three parts: a concave uprise at low energies which
vanishes for zero frequency, a singularity at intermediate energies and a 
convex tail for higher frequencies. It is very interesting to see that all 
these features have  precursors at finite dimerization which are 
captured in the triplon picture.

In the limit of vanishing dimerization the system becomes gapless and the 
spectra therefore start at zero energy. As long as there is some finite 
dimerization the bound state $S_1$ exists and produces the concave uprise at 
small energies and the singularity at intermediate energies resulting from the 
maximum at $k=\pi/2$. We expect that for vanishing dimerization ($\lambda=1$) the 
dispersion of $S_1$ coincides with the lower band edge of the two-triplon continuum 
leading to a square root divergence at the lower band edge for all momenta.
Since the dispersions and the band edges display an extremum at $k=\pi/2$
this divergence leads to the singularity discernible at intermediate energies.
The convex tail at the upper band edge is equally present 
even for strongly dimerized chains, see inset in Fig.~\ref{fig:IR_S0_NN}a. 
It is a consequence of the convex square root behavior at the upper edge of the 
two-triplon continuum.

In Figs.~\ref{fig:IR_S0_NN}b and \ref{fig:IR_S0_NN}c the optical absorption 
at finite frustration $\alpha=0.25$ and $\alpha=0.5$ for the same value of 
dimerization is shown. As discussed earlier the frustration enhances the 
triplon-triplon interaction and increases the binding energy of the 
two-triplon bound states. As can be clearly seen in 
Figs.~\ref{fig:BS_D_S0_l03_NN}b-c and Figs.~\ref{fig:BS_D_S0_l06_NN}b-c the 
spectral weight of $S_1$ at $k=\pi$ increases compared to the weight at $k=\pi/2$. 
Therefore, the first peak in the optical absorption becomes more and more 
prominent on increasing frustration. This leads to the most important 
features at large frustration. At $\alpha=0.5$ the spectral weight of 
$S_2$ is also sizable. Besides the contribution of the two-triplon continuum 
(inset of Fig.~\ref{fig:IR_S0_NN}c) an additional peak appearing for decreasing 
dimerization can be discerned.

The optical absorption for $R^{S=0}_{\rm NNN}$ is plotted in 
Fig.~\ref{fig:IR_S0_NNN}. In the insets an enlargement of the line shapes is 
depicted in order to highlight fine structures. The main difference to the discussion 
of the optical absorption produced by $R_{\rm NN,weak}$ are the consequences 
of the different symmetries of the observables. $R_{\rm NNN}$ suppresses the 
spectral weight for large momenta. Thus the optical response is weak due to 
the phonon form factor which stresses large momenta. In addition, the van-Hove
 singularity resulting from $k=\pi$ of $S_1$ is suppressed so that only a 
weak shoulder can be observed at low energies, independent of frustration and
dimerization. At $\alpha=0.25$ the additional side structures are produced by
the bound states $S_2$ and $S_3$.

\section{Summary}

In this work we have presented results for the spectral densities of the 
dimerized and frustrated Heisenberg chain. We used a perturbative realization 
of the continuous unitary transformations starting from the limit of isolated 
dimers. By the transformations an effective model is obtained which conserves 
the number of triplons. Results for the one-triplon and the 
two-triplon contribution
to the spectral density have been shown for strong ($\lambda=0.3$) and 
intermediate ($\lambda=0.6$) dimerization and for various values of the 
frustration ($\alpha=0; 0.25; 0.5$) and for total spin one and zero.

In the first part of this paper we examined the dynamical structure factor 
which is relevant for inelastic neutron scattering experiments. The 
one-triplon contribution contains most of the spectral weight at strong and 
intermediate dimerization. We provided results for the one-triplon 
dispersion $\omega (k)$ and the $k$-resolved spectral weight $I_1(k)$.
The one-triplon dispersion becomes larger on lowering the dimerization while 
it becomes flatter on increasing the frustration. The spectral weight $I_1(k)$
is mainly concentrated at $k=\pi$. In the limit $\lambda\rightarrow 1$ the 
one-triplon contribution vanishes except for $\alpha=0.5$ around $k=\pi/2$.

Subsequently we discussed the two-triplon contribution to the dynamical 
structure factor. We have provided results for the spectral density of the 
two-triplon continuum and for the dispersion and the spectral weight of the 
two-triplon bound states.

For the unfrustrated spin chain, the spectral weight is concentrated at the 
lower band edge at larger momenta. Two-triplon bound states only exist in a 
finite region about $k=\pi/2$. Increasing the frustration leads to a shift of
spectral weight to higher energies. At $\lambda=0.6$ and $\alpha=0.5$, the 
spectral weight is shifted almost totally to the upper band edge at $k=\pi$. 
This transfer of spectral weight is also found for $\lambda=1$ by exact 
diagonalization at finite temperatures\cite{fabri98b}.

The behavior of the lower band edge changes strongly by the variation of the 
frustration. Generically, we find a square root behavior of the lower band 
edge. It is a consequence of the hardcore interaction between the triplons
which makes it difficult for them to pass each other.
In contrast to this finding, we find a square root divergence at the lower 
band edge for $\alpha=0.25$. Here the 
energy of a two-triplon bound state is degenerate (to the precision of our
analysis) with the lower band  edge of the two-triplon continuum.

We compared the latter finding with results obtained from field 
theory\cite{gogol98}. 
In contrast to our finding for the unfrustrated spin chain,
field theory predicts a square root divergence for the lower band 
edge of the dynamical structure factor of the sine-Gordon model.
Thus the commonly used reduction of the spin
chain to a sine-Gordon by neglecting the marginal operator
cannot be justified quantitatively, at least not for the values of dimerization considered here.
We showed that in the self-consistent harmonic approximation (corresponding to renormalization in first order) the marginal operator $D\cos(4\Phi)$ is as important as
the mass operator $\delta \cos(2\Phi)$ at any finite dimerization.
Renormalization in second order, however, predicts a slow logarithmic suppression 
of the marginal term\cite{affle89} below the critical frustration.

We find that square root behavior represents the generic behavior. A 
square root divergence occurs if a two-triplon bound state is degenerate 
with the lower band edge of the continuum. In the field theoretic language 
this degenerate bound state, which has not yet emerged from the continuum,
is the third breather. We find the concomitant square root divergence for 
$\alpha\approx\alpha_{0,c}$. 

The applicability of the sine-Gordon model to the unfrustrated, but dimerized
 spin chain is further questioned by the study of the excitation energies of 
the bound states. The ratio of the excitation energies of the $S=0$ 
two-triplon bound state and the one-triplon gap is exactly $\sqrt{3}$ in the 
SU(2) -symmetric sine-Gordon model. We find this ratio only for 
$\alpha_0=\alpha_{0,c}$ in agreement with a previous numerical study
\cite{bouze98a}. At present,  we do not know why $\alpha_0=\alpha_{0,c}$
is required to retrieve the field theory result for the second breather,
but $\alpha\approx\alpha_{0,c}$ to retrieve the field theory result for the 
third breather. We cannot exclude completely  
that one has to go to much lower values of
the dimerization, i.e., closer to $\lambda=1$.
But in view of the fact that the marginal operator cannot be 
neglected for a quantitative description, we presume that 
the effective low-energy model to be considered is the double sine-Gordon
model for which the SU(2) symmetry condition will be different
from $K=1/2$ and  hence also the ratios of the breather energies 
will differ from the ratios known for the sine-Gordon model \cite{dashe75b}.
         
In the second part of this work, we discussed spectral properties of the 
dimerized and frustrated spin chain for excitations with total spin zero which
are relevant for optical experiments. We presented results for the 
two-triplon contribution which contains most of the spectral weight.

We examined two different observables: a nearest neighbor (NN) coupling
on the weak bonds $R^{S=0}_{\rm NN,weak}$ and a next-nearest neighbor (NNN)
coupling $R^{S=0}_{\rm NNN}$. The observables obey different symmetries. The
nearest neighbor coupling possesses a reflection symmetry about $k=\pi/2$. 
The next-nearest coupling does not have any reflection symmetry except for
$k=\pi$ where it is odd so that no even $S=0$ two-triplon state can be 
excited. The spectral densities for both observables are dominated by the 
two-triplon bound state $S_1$ which contains most of the spectral weight. This
bound state exists for all momenta independent of dimerization and 
frustration.  The binding energy increases by turning on the frustration.

The spectral weight of the two-triplon continuum is concentrated at the lower 
band edge for all considered values of  dimerization and frustration. 
For finite dimerization,
the lower band edge displays a square root behavior for $\alpha=0.25$
in accordance with the results of the sine-Gordon model.
Divergences may occur only at exactly zero dimerization.
 The 
behavior changes similarly to the $S=1$ case if a two-triplon bound state is 
almost degenerate with the lower band edge of the two-triplon continuum.
Such a degeneracy appeared in our data for $\alpha\approx 0$ and 
for $\alpha\approx 0.5$.
      
Finally, we presented results for the Raman response and the 
infrared absorption. Both experiments are dominated for the values of
dimerization considered  by the bound state $S_1$. 
This bound state produces two van-Hove
singularities in the infrared absorption resulting from $k=\pi/2$ and 
$k=\pi$. The van-Hove singularity at lower energies becomes more important
for larger values of the frustration.
     
\section{Conclusions} 

We have shown that continuous unitary transformations (CUTs) are an excellent 
tool to calculate spectral densities in high resolution for the dimerized and 
frustrated Heisenberg chain. We provided a detailed study of the 
spectral densities and extracted the generic features of the spectral 
properties. These data will help to analyze a large variety of spectroscopic
measurements for dimerized and frustrated spin chains systems.

We used a description in terms of triplons (elementary triplets) carrying 
$S=1$. Previously\cite{schmi03c}, we had shown for unfrustrated 
chains that a description in terms of two triplons is sufficient even in the 
limit of zero dimerization. The triplon may also serve as an elementary
excitation of the uniform Heisenberg chain besides the well-established 
spinon excitation. In the present work, we found further strong support for this 
result. The two-triplon spectral densities computed at finite dimerization 
displayed well-developed precursors of the results for the uniform chain 
based on spinons, e.g., the dynamical structure factor probing the $S=1$ 
sector\cite{karba97} or the 
optical absorption\cite{loren97c} probing the $S=0$ sector.

In the frustrated case it is as yet an open issue whether 
a triplonic description works also in the limit of zero dimerization. 
Especially the gapped phase $(\alpha_0>\alpha_{0,c})$ requires that
the two-spinon continuum between $\Delta$ and $2\Delta$ can be described
by a dense distribution of bound many-triplon states. For this to occur an
infinite-range effective interaction is necessary which is beyond
the scope of the perturbative CUTs.
But we take the nice agreement between the shifts of spectral weight obtained 
by  
complete exact diagonalization at finite temperatures\cite{fabri98b} for the 
undimerized frustrated chain with our results at finite dimerization as
indication that the triplon description can be extended to the undimerized frustrated chain, too.

The comparison of our results to those obtained by mapping the spin chain
to a sine-Gordon model led to a number of insights. Both approaches 
agree that the generic singularity at the lower band edge is a square root,
either a divergence or a zero. The divergence occurs if and only if a bound
state is degenerate with the band edge. It turned out that the predictions of
the single sine-Gordon model hold for critical frustration only in agreement
with previous conclusions 
based on numerical results\cite{bouze98a}. 
The general
spin chain at non-vanishing dimerization requires to go beyond the single sine-Gordon model. 
We showed
analytically that in the self-consistent harmonic approximation
(equivalent to first order renormalization)
the Umklapp term is as large as the mass term
\cite{notiz4}.

The present study based on perturbative continuous unitary transformation is
limited by the tractable maximum range of hopping and interaction 
processes which correlates with the maximum order. Thus an investigation 
of the spectral properties for small and zero gaps, i.e., dimerizations, 
will require to develop new methods allowing for
larger ranges (correlation lengths). The continuous unitary transformations
do not need to be realized perturbatively. Ongoing research\cite{reisc03a} 
deals with a self-similar realization which renders the treatment of larger
ranges possible. In this way, a closer look at critical systems
and systems with massive spinons will come within reach.

\begin{acknowledgments}
We thank A.~M.~Tsvelik, I.~ Affleck, A.~Reischl, A.~G\"o\ss ling, M.~Gr\"uninger and 
E.~M\"uller-Hartmann for stimulating and encouraging discussions and the
 DFG  for financial support in SP 1073 and in SFB 608. 
\end{acknowledgments}


\begin{thebibliography}{10}

\bibitem{halda82b}
F.~D.~M. Haldane, Phys. Rev. B {\bf 25},  4925  (1982).

\bibitem{cloiz62}
J. des Cloizeaux and J.~J. Pearson, Phys. Rev. {\bf 128},  2131  (1962).

\bibitem{fadde81}
L.~D. Faddeev and L.~A. Takhtajan, Phys. Lett. {\bf 85A},  375  (1981).
\bibitem{karba97}
M. Karbach, G. M\"uller, A.~H. Bougourzi,A. Fledderjohann and K. M\"utter , Phys. Rev. B {\bf 55},  12510  (1997).
\bibitem{knett01b}
C. Knetter, K.~P. Schmidt, M. Gr\"uninger, and G.~S. Uhrig, Phys. Rev. Lett.
  {\bf 87},  167204  (2001).

\bibitem{schmi01}
K.~P. Schmidt, C. Knetter, and G.~S. Uhrig, Europhys. Lett. {\bf 56},  877
  (2001).

\bibitem{zheng03a}
W. Zheng, C.~J. Hamer, and R.~R.~P. Singh, Phys. Rev. Lett. {\bf 91}, 037206 (2003).
\bibitem{hamer03}
C.~J. Hamer, W. Zheng, and R.~R.~P. Singh, Phys. Rev. B {\bf 68}, 214408 (2003)
\bibitem{hase93a}
M. Hase, I. Terasaki, and K. Uchinokura, Phys. Rev. Lett. {\bf 70},  3651
  (1993).

\bibitem{nishi94}
M. Nishi, O. Fujita, and J. Akimitsu, Phys. Rev. B {\bf 50},  6508  (1994).

\bibitem{riera95}
J. Riera and A. Dobry, Phys. Rev. B {\bf 51},  16098  (1995).

\bibitem{casti95}
G. Castilla, S. Chakravarty, and V.~J. Emery, Phys. Rev. Lett. {\bf 75},  1823
  (1995).

\bibitem{isobe96}
M. Isobe and Y. Ueda, J. Phys. Soc. Jpn. {\bf 65},  1178  (1996).

\bibitem{garre97a}
A.~W. Garrett, S.~E. Nagler, D.~A. Tennant, B.~C. Sales, and T. Barnes, Phys. Rev. Lett. {\bf 79}, 745 (1997).

\bibitem{chabo97a}
G. Chaboussant, P.~A. Crowell, L.~P. L\'{e}vy, O. Piovesana, A. Madouri and D. Mailly, Phys. Rev. B {\bf 55},  3046  (1997).

\bibitem{xu00}
G. Xu, C. Broholm, D.~H. Reich, and M.~A. Adams, Phys. Rev. Lett. {\bf 84},
  4465  (2000).

\bibitem{tenna03}
D.~A. Tennant, C. Broholm, D.~H. Reich, S.~E. Nagler, G.~E. Granroth, T. Barnes, K. Damle, G. Xu, Y. Chen and B.~C. Sales, Phys. Rev. B {\bf 67},  054414  (2003).

\bibitem{nagle91}
S.~E. Nagler, D.~A. Tennant, R.~A. Cowley, T.~G. Perring and  S.~K. Satija , Phys. Rev. B {\bf 44},  12361  (1991).

\bibitem{ami95}
T. Ami, M.~K. Crawford, R.~L. Harlow, Z.~R. Wang, D.~C. Johnston, Q. Huang, R.~W. Erwin, Phys. Rev. B {\bf 51},  5994  (1995).

\bibitem{motoy96}
N. Motoyama, H. Eisaki, and S. Uchida, Phys. Rev. Lett. {\bf 76},  3212
  (1996).

\bibitem{notiz1}
Previously, we used the term 
``elementary triplets'' to distinguish 
three-fold degenerate elementary excitations from
magnons which are the elementary excitations
of long-range \emph{ordered} magnets.  We henceforth 
use the term ``triplon'' instead to have a
shorter expression distinguishing  more clearly
from composite triplets.

\bibitem{schmi03c}
K.~P. Schmidt and G.~S. Uhrig, Phys. Rev. Lett. {\bf 90},  227204  (2003).

\bibitem{julli83}
R. Jullien and F.~D.~M. Haldane, Bull. Am. Phys. Soc. {\bf 28},  344  (1983).

\bibitem{okamo92}
K. Okamoto and K. Nomura, Phys. Lett. {\bf A169},  433  (1992).

\bibitem{egger96}
S. Eggert, Phys. Rev. B {\bf 54},  R9612  (1996).

\bibitem{majum69a}
C.~K. Majumdar and D.~K. Ghosh, J. Math. Phys. {\bf 10},  1388  (1969).

\bibitem{majum69b}
C.~K. Majumdar and D.~K. Ghosh, J. Math. Phys. {\bf 10},  1399  (1969).

\bibitem{broek80}
P.~M. van~den Broek, Phys. Lett. {\bf 77A},  261  (1980).

\bibitem{shast81a}
B.~S. Shastry and B. Sutherland, Phys. Rev. Lett. {\bf 47},  964  (1981).

\bibitem{chitr95}
R. Chitra, Swapan Pati, H.~R. Krishnamurthy, Diptiman Sen and S. Ramasesha, Phys. Rev. B {\bf 52},  6581  (1995).

\bibitem{luthe75}
A. Luther and I. Peschel, Phys. Rev. B {\bf 12},  3908  (1975).

\bibitem{halda80}
F.~D.~M. Haldane, Phys. Rev. Lett. {\bf 45},  1358  (1980).

\bibitem{uhrig96b}
G.~S. Uhrig and H.~J. Schulz, Phys. Rev. B {\bf 54},  R9624  (1996).

\bibitem{gogol98}
A.~O. Gogolin, A.~A. Nersesyan, and A.~M. Tsvelik, {\em Bosonization and
  Strongly Correlated Systems} (Cambridge University Press, Cambridge, UK,
  1998).

\bibitem{affle97}
I. Affleck,  in {\em Dynamical Properties of Unconventional Magnetic Systems},
  Vol.~349 of {\em NATO SCIENCE SERIES E: Applied Sciences}, edited by A.~T.
  Skjeltorp and D. Sherrington (Kluwer Academic Publishers, Dordrecht, 1998).

\bibitem{fledd97}
A. Fledderjohann and C. Gros, Europhys. Lett. {\bf 37},  189  (1997).

\bibitem{bouze98a}
G. Bouzerar, A.~P. Kampf, and G.~I. Japaridze, Phys. Rev. B {\bf 58},  3117
  (1998).

\bibitem{shevc99}
P. Shevchenko, V.~N. Kotov, and O.~P. Sushkov, Phys. Rev. B {\bf 60},  3309
  (1999).

\bibitem{barne99}
T. Barnes, J. Riera and D.~A. Tennant, Phys. Rev. B {\bf 59}, 11384 (1999).

\bibitem{trebs00}
S. Trebst, H. Monien, C.~J. Hamer, Z. Weihong and R.~R.~P. Singh, Phys. Rev. Lett. {\bf 85},  4373  (2000).

\bibitem{zheng01a}
W. Zheng, C.~J. Hamer, R.~R.~P. Singh, S. Trebst and H. Monien, Phys. Rev. B {\bf 63},  144410 and 144411  (2001).

\bibitem{wegne94}
F.~J. Wegner, Ann. Physik {\bf 3},  77  (1994).

\bibitem{knett00a}
C. Knetter and G.~S. Uhrig, Eur. Phys. J. B {\bf 13},  209  (2000).

\bibitem{home}
www.thp.uni-koeln.de/~gu

\bibitem{knett03b}
C. Knetter, K.~P. Schmidt, and G.~S. Uhrig, Eur. Phys. J. B {\bf 36}, 525 (2004).

\bibitem{knett03a}
C. Knetter, K.~P. Schmidt, and G.~S. Uhrig, J. Phys.: Condens. Matter {\bf 36},
   7889  (2003).

\bibitem{schmi03a}
K.~P. Schmidt, C. Knetter, and G.~S. Uhrig, Acta Physica Polonica B {\bf 34},
  1481  (2003).

\bibitem{schmi03d}
K.~P. Schmidt, H. Monien, and G.~S. Uhrig, Phys. Rev. B {\bf 67},  184413
  (2003).

\bibitem{notiz2}
Even the one-triplon contribution does not
vanish completely since there are regions in momentum space 
where the one-triplon state is stable even at zero dimerization
(see also discussion at the end of IV.A).

\bibitem{mulle03}
M. M\"uller and H.-J. Mikeska,   (2003).

\bibitem{fabri98b}
K. Fabricius and U. L\"ow, Phys. Rev. B {\bf 57},  13371  (1998).

\bibitem{caspe82}
W.~J. Caspers and W. Magnus, Phys. Lett. {\bf 88A},  103  (1982).

\bibitem{caspe84}
W.~J. Caspers, K.~M. Emmett, and W. Magnus, J. Phys. A: Math. Gen. {\bf 17},
  2687  (1984).

\bibitem{singh99a}
R.~R.~P. Singh and Z. Weihong, Phys. Rev. B {\bf 59},  9911  (1999).

\bibitem{smirn92}
F.~A. Smirnov, {\em Form Factors in Completely Integrable Models of Quantum
  Field Theory} (World Scientific, Singapore, 1992).

\bibitem{affle86b}
I. Affleck, Nucl. Phys. B {\bf 265},  409  (1986).

\bibitem{affle87}
I. Affleck and F.~D.~M. Haldane, Phys. Rev. B {\bf 36},  5291  (1987).

\bibitem{white96}
S.~R. White and I. Affleck, Phys. Rev. B {\bf 54},  9862  (1996).

\bibitem{chitr97}
R. Chitra and T. Giamarchi, Phys. Rev. B {\bf 55},  5816  (1997).

\bibitem{dashe75b}
R.~F. Dashen, B. Hasslacher, and A. Neveu, Phys. Rev. D {\bf 12},  2443
  (1975).

\bibitem{nakan80}
T. Nakano and H. Fukuyama, J. Phys. Soc. Jpn. {\bf 49},  1679  (1980).

\bibitem{nakan81}
T. Nakano and H. Fukuyama, J. Phys. Soc. Jpn. {\bf 50},  2489  (1981).

\bibitem{cross79}
M.~C. Cross and D.~S. Fisher, Phys. Rev. B {\bf 19},  402  (1979).

\bibitem{affle89}
I. Affleck, D. Gepner, J.~J. Schulz, and T. Ziman, J. Phys. A {\bf 22}, 511 (1989); {\bf 23}, 4725 (1990).
\bibitem{notiz4}
G.~S.~Uhrig,~Habilitation 1999,~www.thp.Uni-Koeln.DE/~gu/veroeffentlichungen.html

\bibitem{notiz3}
E.~M\"uller-Hartman and G.S. Uhrig, unpublished.

\bibitem{zheng01b}
W. Zheng, C.~J. Hamer, R.~R.~P. Singh, S. Trebst and H. Monien, Phys. Rev. B {\bf 63},  144411  (2001).

\bibitem{reisc03a}
A. Reischl, E. M\"uller-Hartmann, and G.~S. Uhrig, cond-mat/0401028 (unpublished).

\bibitem{loren95a}
J. Lorenzana and G.~A. Sawatzky, Phys. Rev. Lett. {\bf 74},  1867  (1995).

\bibitem{loren95b}
J. Lorenzana and G.~A. Sawatzky, Phys. Rev. B {\bf 52},  9576  (1995).

\bibitem{fleur68}
P.~A. Fleury and R. Loudon, Phys. Rev. {\bf 166},  514  (1968).

\bibitem{shast90b}
B.~S. Shastry and B.~I. Shraiman, Phys. Rev. Lett. {\bf 65},  1068  (1990).

\bibitem{windt01}
M. Windt, M. Gr\"uninger, T. Nunner, C. Knetter, K.~P. Schmidt,3 G.~S. Uhrig, T. Kopp, A. Freimuth, U. Ammerahl, B. B\"uchner and A. Revcolevschi, Phys. Rev. Lett. {\bf 87}, 127002 (2001)

\bibitem{loren97c}
J. Lorenzana and R. Eder, Phys. Rev. B {\bf 55},  R3358  (1997).

\bibitem{suzuu96}
H. Suzuura {\it et~al.}, Phys. Rev. Lett. {\bf 76},  2579  (1996).

\end{thebibliography}
\end{document}